\documentclass[trans]{IEEEtran}
\usepackage{hyperref}
\hypersetup{
colorlinks=true,
urlcolor=blue,
citecolor=blue}

\usepackage{cite}
\usepackage{amsmath,amssymb,amsfonts}
\usepackage{algorithmic}
\usepackage{graphicx}
\usepackage{textcomp}
\usepackage{wrapfig}

\usepackage{multicol}

\usepackage{ltablex}
\usepackage{tabu}
\usepackage{adjustbox}

\usepackage[all]{xy,xypic}
\usepackage{amsfonts,amssymb,amsmath,amsgen,amsopn,amsbsy,theorem,graphicx,epsfig}
\usepackage{eufrak,amscd,bezier,latexsym,mathrsfs,enumerate}\usepackage[utf8]{inputenc}\usepackage[english]{babel}
\usepackage[dvipsnames]{xcolor}
\usepackage[pagewise]{lineno}
\usepackage{amssymb}
\usepackage{amsmath}
\usepackage{multicol}
\usepackage{enumitem}
\usepackage{subfig}
\usepackage{multirow}
\usepackage{setspace} 
\usepackage{lipsum}
\usepackage{adjustbox}

\usepackage{optidef}

\usepackage{graphicx} 
\graphicspath{{figs/}} 
\usepackage{subfig}
\usepackage{multirow}
\usepackage{color,soul}

\begin{document}
\title{Adaptive Control of IoT/M2M Devices in Smart Buildings using Heterogeneous Wireless Networks}
\author{Rania Djehaiche, Salih Aidel, Ahmad Sawalmeh, Nasir Saeed \IEEEmembership{Senior Member, IEEE,} and Ali H. Alenezi
\thanks{Rania Djehaiche is with the ETA Laboratory, Department of Electronics, Faculty of Science and Technology,  University Mohamed El Bachir El Ibrahimi of Bordj Bou Arreridj, 34030, Algeria, (e-mail: \texttt{rania.djehaiche@univ-bba.dz}). }
\thanks{Salih Aidel is with the 
 Department of Electronics, Faculty of Science and Technology, University Mohamed El Bachir El Ibrahimi of Bordj Bou Arreridj, 34030, Algeria, (e-mail: \texttt{salih.aidel@univ-bba.dz}).}
 \thanks{Ahmad~Sawalmeh is with the Data Science and Artificial Intelligence Department - College of Science and Information Technology, Irbid National University, Irbid, Jordan, (e-mail: \texttt{a.sawalmeh@inu.edu.jo}).}
\thanks{Nasir Saeed is with the
Department of Electrical and Communication Engineering, United Arab Emirates  University (UAEU), Al Ain, UAE, (e-mail: \texttt{mr.nasir.saeed@ieee.org).}}
\thanks{Ali H. Alenezi is with the
Department of Electrical Engineering, Northern Border University, Arar, 
 Saudi Arabia, (e-mail: \texttt{ali.hamdan@nbu.edu.sa).}\\ (Corresponding author: Nasir Saeed.)}}

\maketitle
\begin{abstract}
With the rapid development of wireless communication technology, the Internet of Things (IoT) and Machine-to-Machine (M2M) are becoming essential for many applications. One of the most emblematic IoT/M2M applications is smart buildings. The current Building Automation Systems (BAS) are limited by many factors, including the lack of integration of IoT and M2M technologies, unfriendly user interfacing, high costs, using no more than one or two wireless communication networks, limited wireless transmission range, and the lack of a convergent solution. Therefore, this paper proposes a better approach of using heterogeneous wireless networks consisting of Wireless Sensor Networks (WSNs) and Mobile Cellular Networks (MCNs) for IoT/M2M smart building systems. {The proposed system is an inexpensive embedded system that comprises Arduino and NodeMCU} boards with several compatible sensors, actuators, and modules for controlling and collecting data over heterogeneous communication technologies (RFID, Bluetooth, Wi-Fi, GSM, LTE). { All collected data is uploaded to the ThingSpeak platform, allowing the building system to be monitored via the ThingSpeak webpage or the Raniso app.  One of the most significant outcomes of this research is to provide accurate readings to the server, and very low latency, through which users can easily control and monitor remotely the proposed system that consists of several innovative services, namely smart parking, garden irrigation automation, intrusion alarm, smart door, fire and gas detection, smart lighting, smart medication reminder, and indoor air quality monitoring.} All these services are designed and implemented to control and monitor from afar the building via our free mobile application named “Raniso” which is a local server that allows remote control of the building via RFID/Bluetooth/Wi-Fi connectivity and cellular networks. {This IoT/M2M smart building system is customizable to meet the needs of users, improving safety and quality of life while reducing energy consumption. Additionally, it helps prevent the loss of resources and human lives by detecting and managing risks.}
\end{abstract}


\section{Introduction}
\label{sec:introduction}
\IEEEPARstart{T}he Internet of Things (IoT) and Machine-to-Machine (M2M) are the fastest growing areas of cutting-edge technologies, with the number of connected devices now massively outnumbering humans. The International Telecommunication Union (ITU) defines the IoT as a global infrastructure for the information society, enabling advanced services by interconnecting physical and virtual things based on existing and evolving interoperable information and communication technologies~\cite{1}. M2M is considered an integral part of the IoT ecosystem, defined as a communication that allows devices to connect with one another across a wired or wireless communication network without the need for human intervention ~\cite{3}. Intelligent sensors, advanced wireless technologies, and autonomic computing software are used to assist a network device in understanding and transmitting data and making decisions in an M2M system \cite{3_n}. Recently, the evolution of numerous wireless communication technologies, such as intelligent reflecting surfaces \cite{Kunrui2022}, massive MIMO \cite{Wang2022},  and ambient backscattering \cite{Zhang2019} have substantially increased the potential capabilities of IoT/M2M technologies and made them more common than ever, as their convergence has become necessary to meet the ever-increasing requirements of IoT/M2M. { Among different wireless communication technologies, ambient backscatter has been emerging for large-scale IoT/M2M applications that can allow transmission of data without requiring power from the IoT device and facilitates device-to-device (D2D) and even multi-hop communications. By reflecting far-field electromagnetic (EM) waves from ambient radio frequency (RF) transmitters, a tiny passive gadget can send data with very low power consumption. This cutting-edge technology with can effectively solve the battery problem faced by massive low-power devices in large-scale IoT/M2M communications \cite{d2017, Yao2020, s22197260}.} The heterogeneous networks {(HetNets)} operating on different wireless communication technologies, especially  wireless sensor networks (WSNs) and mobile cellular networks (MCNs), offer numerous applications in many industrial sectors such as home automation, industry, healthcare, agriculture, and smart cities~\cite{5}.

Among these applications,  smart buildings are emerging, which support the flow of information throughout the building, providing advanced functionalities and services, allowing the automatic control, monitoring, management, and maintenance of the various subsystems or applications of the building in an optimal and integrated way, locally and/or remotely ~\cite{7}.  
{There are many research papers dealing with wireless network-based IoT/M2M smart building systems, whether MCN or WSN } \cite{6_n,7_n,8_n,26,21,11_n,12_n,13_n,14_n}, {which will be discussed in detail in the related work part. Comparing the proposed system with all these research works,}
the drawbacks of existing building management systems include the lack of using IoT and M2M technologies together, reliance on one or two wireless communication technologies at most, a limited wireless transmission range, mostly inconveniently designed user interface, and excessive costs. Therefore, our approach presents a hybrid (local and remote) and a low-cost IoT/M2M-based smart building system with a user-pleasant smartphone interface. The proposed system comprises several main services: smart parking, garden irrigation automation, intrusion alarm, smart door, fire and gas detection, smart lighting, smart medication reminder, and indoor air quality monitoring. 
All these services can be controlled and monitored remotely by {our mobile application named} ‘Raniso’, which is used as a local server to control the building via different {HetNets} such as RFID/Bluetooth/WiFi connectivity and cellular networks like GSM, 4G, or 5G. In addition, the proposed system connects all smart devices in an energy-efficient, secure, convenient, and cost-effective manner. 
This paper is an extension of our previous works in  \cite{7,9} to develop an IoT/M2M smart building prototype using several compatible sensors, actuators, and modules, besides the { Raniso App. In addition to receiving data from the smart building, the system also stores data in the cloud database and displays it visually on the ThingSpeak webpage, as well as remotely monitoring it via the Raniso App.}    


\section{ STATE OF THE ART}
\subsection{Motivations and Research Contributions}
{Heterogeneous wireless networks are of growing interest in several areas, including IoT/M2M smart building automation which has developed at a rapid pace over the past few years. The main benefits of smart buildings are reduced maintenance expenses, reduced energy consumption, comfort, tranquility, entertainment, safety, security, improved productivity, more livable structures, and higher resale value. Despite the great applicability of the IoT in the implementation of smart building systems, there is a lack of strategies that involve the integration of IoT and M2M technologies in most existing systems. Besides relying on only one or two wireless communication technologies, most of the existing studies also suffer from limited wireless range connectivity, poorly designed user interfaces, and high costs.
As a means of addressing these issues and reducing existing smart building system limitations, this study proposes an adaptive control of IoT/M2M devices in smart buildings using heterogeneous wireless networks to expand the scope of connectivity and allow users to control their building easily and efficiently by choosing the appropriate connection, whether it is WSN or MCN, through a user-friendly interface using smartphones via the Raniso App, regardless of time and location. Costs are reduced by using effective and affordable components, as well as the Raniso App, which is free to manage, monitor, and control building devices and conditions over various heterogeneous networks.
The research contribution is summarized as a design for adaptive control of IoT/M2M devices in smart buildings using our developed mobile application Raniso via heterogeneous wireless networks. As a result, we have listed the contributions of this research work below:}

\begin{itemize}
    \item {Discussing and comparing related work on IoT/M2M smart building-based wireless technologies with the proposed system.}
    \item {Development, implementation, and design of adaptive control for IoT/M2M devices connected to heterogeneous networks in smart building systems using Arduino and NodeMCU boards with heterogeneous low-cost and small-sized sensors and actuators.}
	\item {Dynamic use and intelligent management of IoT/M2M devices to facilitate monitoring and control of smart building systems via IoT platform ThingSpeak and via our free mobile app "Raniso".}
    \item {Proposing eight innovative services for smart buildings, including smart parking, garden irrigation automation, intrusion alarm, smart door, fire and gas detection, smart lighting, smart medication reminder, and indoor air quality monitoring.}
    \item	{Validation of the functionality of the proposed system in terms of adaptation, control, automation, safety, comfort, energy efficiency, and performance. }
\end{itemize}

\subsection{Related Works}
{Recently, with the evolution of IoT/M2M applications based on heterogeneous wireless networks, there has been a rapid progression of smart home applications with a gradual increase in the size of the automated environments, starting from the living room and moving to the apartment, then to entire buildings, and finally to the more general scenario of smart cities. In this section, previous research works related to IoT/M2M smart building-based wireless technologies, either WSNs or MCNs or both, are reviewed.
Researchers in} \cite{6_n} {present the development of an Artificial Intelligence-Based Smart Building Automation Controller (AIBSBAC) that includes an intelligent user identification subsystem, an intelligent decision-making subsystem, internal and external environment observation subsystems, and a universal infrared communication system. The AIBSBAC system can dynamically adapt to the user's choices, which are aimed at improving energy efficiency, user comfort, and security. Heterogeneous wireless connectivity, represented by RF, Ethernet, and Bluetooth technologies, connects AIBSBAC to electrical devices. In contrast, the proposed system only included a few applications and ignored many of the other important applications (such as fire detection and gas detection). The work in} \cite{7_n}, { presents a smart building system adaptable to office environments. The proposed system is based on the   ZigBee wireless sensor network has been coupled with Java and Android; to develop a platforme for  
automated and manual control of devices to track real-time environmental data of smart building systems.  However, using only ZigBee as a WSN in the system has disadvantages, including short range, low data transmission speed, high maintenance cost, low transmission, as well as low network stability.}

In \cite{8_n}, {the authors focused on monitoring home appliances via IoT using various sensors to monitor temperature, fire, and gas, and an LCD screen to display the sensor's values. When gas leaks or fires are discovered, the system quickly sends a text message to the user's cell phone, activates the siren, and the spray engine is activated and displays a message on an LCD screen to warn users. While the IoT is used to improve security, where communication between sensors and transducers is accomplished wirelessly using a single chip via Wi-Fi. Despite this, the system has some limitations: The WiFi data transfer rate decreases as the number of users increases, and GSM is expensive due to SMS charges. }

{A home prototype system based on heterogeneous wireless networks (Wi-Fi, Bluetooth, and RFID) for control and monitoring is proposed in} \cite{26}.{ The system is based on two components:  the first one is an automation system that was built using an Arduino UNO, which is responsible for reading and processing different types of sensor values. The second component is the security system and outdoor lighting using NodeMCU to monitor the home security status from anywhere through a specialized Graphical User Interface (GUI) that was programmed in HTML and allows the user to monitor the home security and turn ON and OFF the outdoor lights using a particular internet protocol (IP) address provided by NodeMCU. However, the applications implemented are very basic and not innovative, in addition to neglecting the user's right to choose their wireless communication preferences.}

In \cite{21}, {an IoT-based portable automation system called IoT@HoMe for smart homes was designed and manufactured using NodeMCU as the microcontroller and Internet gateway. In addition to using several sensors to monitor various parameters related to the building, various actuators perform control activities of home devices. The main objective of this research is to create an efficient, affordable, and portable system that can easily monitor home conditions and operate home devices over the Internet at any time and from any location. Similarly, another research work} \cite{11_n} {presented a multifunctional smart home system using a micro-web server based on an Arduino Yún microcontroller with Internet connectivity via an Android-based mobile application. The system can be controlled even if there is no WiFi connection available, as it can be accessed via MCNs (e.g. 3G or 4G networks). Despite this, both of these researches }\cite{21,11_n} {remain limited in that they rely solely on the Internet for their communications and did not take into account the unavailability or interruptions of the Internet in some regions.}

In \cite{12_n}, { an IoT-based smart building system for indoor environmental management was proposed and implemented using a Raspberry Pi 3 with a camera, an obstacle sensor, and several environmental sensors to detect smart building parameters such as temperature, humidity, brightness, and air quality. The Raspberry Pi was programmed to collect data from the sensors and a photo taken by the camera every 15 minutes. In addition, the Raspberry Pi was equipped with a light neural network that could count the number of people in the room based on the photos taken by the camera. When the user is outside the coverage area of the Wi-Fi AP, he is unable to contact the server and cannot directly use the smartphone to transmit commands to the Raspberry Pi controller.}
 {Researchers in} \cite{13_n}  {discuss an affordable, safe, and energy-efficient WiFi-based smart home system that enables homeowners to monitor their appliances from their mobile phones at home or remotely. This system uses Raspberry Pi and Arduino Mega as microcontrollers and internet gateways for IoT automation monitoring and control. Numerous sensors and actuators are used in the system to monitor multiple home-related parameters via the Blynk app. Despite this, sometimes Blynk App becomes unresponsive and displays data/output that could be false in some cases. Also, in some cases, Blynk App's notifications are delayed in sending to the user. }
In \cite{14_n}  { a smart building fire and gas leakage alert system namely SB112, which combines a small-size multisensor-based scheme with an open-source edge computing framework and an automated next-generation (NG) 112 emergency call functionality was proposed using ESP32 as microcontrollers units and Raspberry Pi as Edge gateway. As part of an end-to-end scenario, crucial actors such as IoT devices, public safety answering points (PSAP), middleware for a smart city platform, and relevant operators are involved. However, only the fire and gas leakage alert system was used, while other important systems that should be in the smart building, related to energy consumption and comfort, for example, were ignored.}

Table I  {summarizes and reviews previous research works related to wireless network-based IoT/M2M smart building systems. The comparison of our proposed IoT/M2M smart building using heterogeneous wireless networks with existing systems also highlights its main advantages over these systems. As shown in Table I, the proposed system aims to overcome the limitations of the existing systems. It is based on a simulated scenario with a real implementation for validation purposes. In addition, the main advantages of our system include the ability to connect to the Internet, cloud platforms, and various heterogeneous wireless networks, with support for multiple heterogeneous devices (sensors, actuators, and shields). Users also have the option to control IoT/M2M devices using either push buttons, WSN (Bluetooth or Wi-Fi), MCN (GSM or LTE), or voice commands via Google Assistant. In addition, the Raniso app, which consists of user-friendly graphical interfaces that control and monitor building system devices and alert users in case of danger, as well as a visual display that allows users to monitor all building parameters in real-time without the Internet, both indoors and outdoors, such as temperature, humidity, air quality, building condition, whether it is safe or not, etc. Furthermore, the system we proposed consists of many smart applications and services that should be part of any smart building system including smart parking, garden irrigation automation, intrusion alarm, smart door, fire and gas detection, smart lighting, smart medication reminder, and indoor air quality monitoring.    All of these proposed services contributed to improving safety, comfort, energy savings, and internal and external care in the building system. }

\begin{table*}
 	\scriptsize
 	\renewcommand{\arraystretch}{1}
 	\caption{\uppercase{Comparison of related work on. }}
 	\label{tableint1}
 	\centering
 	\begin{tabular}{|c|c|c|c|c|c|c|c|c|c|c|}
 		\hline 
 		Reference&\cite{6_n}  &\cite{7_n}&\cite{8_n} &\cite{26}&\cite{21} &\cite{11_n} &\cite{12_n}& \cite{13_n}&\cite{14_n}&Proposed \\
 		Num&&&&&&&&&&System \\
 		\hline
 	Smart system	&Building&Building&Home&Home&Home&Home&Building&Home&Building&Building \\
 		\hline
 	Year	&2015&2018&2018&2018&2019&2019&2021&2021&2022&2022 \\
 		\hline
 	Controller &Arduino	&STM32&	PC Server&	Arduino/
&	NodeMCU	&Arduino&	RaspberryPi&	Arduino/&	ESP32/
&	Arduino/
\\
 &&	 &&	NodeMCU&		&&	&	 RaspberryPi&	
Raspberrypi& NodeMCU
\\
 		\hline
 	Protocol	&AIBSBAC&	ZigBee&	Wi-Fi/GSM&	RFID/Wi-Fi&	Wi-Fi&	Wi-Fi/LTE&	Wi-Fi&	Wi-Fi&	Wi-Fi&	RFID/Wi-Fi/
\\
&&	 &&	Bluetooth (BT)&		&&	&	 &	& BT/GSM/LTE
 \\
	\hline
 	WSN	&\checkmark&\checkmark&\checkmark&\checkmark&\checkmark&\checkmark&\checkmark&\checkmark&\checkmark&\checkmark\\
 		\hline
 	MCN	&&&\checkmark&&&\checkmark&&&&\checkmark\\
 		\hline
 		
 	Outdoor Control	&\checkmark&&&&\checkmark&\checkmark&\checkmark&\checkmark&\checkmark&\checkmark\\
 		\hline
 	Indoor Control	&\checkmark&\checkmark&&\checkmark&\checkmark&\checkmark&\checkmark&\checkmark&\checkmark&\checkmark\\
 		\hline
 		
 	Outdoor Care	&&&&\checkmark&&&&&&\checkmark\\
 		\hline
 	Indoor Care	&\checkmark&\checkmark&\checkmark&\checkmark&\checkmark&\checkmark&\checkmark&\checkmark&\checkmark&\checkmark\\
 		\hline
 	
     Safety	&\checkmark&&\checkmark&\checkmark&\checkmark&\checkmark&\checkmark&\checkmark&\checkmark&\checkmark\\
 		\hline
 		
     Energy Efficient	&\checkmark&\checkmark&&&\checkmark&&\checkmark&\checkmark&&\checkmark\\
 		\hline
 		
 	Comfort	&\checkmark&\checkmark&&\checkmark&\checkmark&\checkmark&\checkmark&\checkmark&&\checkmark\\
 		\hline
 		
 	Monitoring	&\checkmark&\checkmark&\checkmark&\checkmark&\checkmark&\checkmark&\checkmark&\checkmark&\checkmark&\checkmark\\
 		\hline
 	
 Smartphone	&\checkmark&&&\checkmark&\checkmark&\checkmark&\checkmark&\checkmark&&\checkmark\\
 		\hline
 	Web-based	&&&&\checkmark&\checkmark&\checkmark&&&&\checkmark\\
 		\hline
 		
 	Google Assistant	&&&&&\checkmark&&&&&\checkmark\\
 		\hline
 			User Preferences	&\checkmark&&&&&&&&&\checkmark\\
 		\hline
 		
 Virtual Simulation	&&\checkmark&&&&&&\checkmark&&\checkmark\\
 of system design	&&&&&&&&&&\\
 		\hline
 		
 	Real Implementation	&\checkmark&&&\checkmark&\checkmark&\checkmark&\checkmark&\checkmark&\checkmark&\checkmark\\
 		\hline
 		
	\end{tabular}
 \end{table*}

\section{Proposed IoT/M2M-based smart building system using heterogeneous networks
}
The proposed heterogeneous network infrastructure for the IoT/M2M smart building consists of wireless sensor networks and mobile-cellular networks. Wireless sensor networks are concerned with communication between devices in the building and include wireless local area networks (WLANs) such as Wi-Fi (IEEE802.11x), which support small area connectivity, and wireless personal area networks (WPANs) like Bluetooth (IEEE 802.15. 1) and RFID, which support short-range connectivity for communication between personal devices. In contrast, mobile cellular networks are the communication network between devices in the building and devices in the outside network and include wireless wide area networks (WWANs) that use mobile telecommunication cellular network technologies such as 2G (GSM), 3G (UMTS), 4G (LTE), and 5G.
{The following expression describes the proposed heterogeneous network infrastructure for the IoT/M2M smart building system.}
\begin{equation}\label{Het}
\begin{split}
       HetNets = WSNs_{WLANs_{WiFi}} + WPANs_{Blutooth,RFID} \\+ MCNs_{WWANs_{GSM,UMTS,LTE}}
       \end{split}
\end{equation}

The convergence of two heterogeneous networks, like WSNs and MCNs, enhances M2M communication and IoT technology. Converged networks can be divided into four categories: device convergence, protocol convergence, service convergence, and full convergence of any user’s devices, communication protocols, and network services \cite{Li2021, 14}. {The full convergence can be expressed as:}
\begin{equation}\label{full}
\begin{split}
       Full_{Convergence} = Device_{Convergence} \\+ Protocol_{Convergence} +Service_{Convergence}
       \end{split}
\end{equation}

In WSNs, the smart mobile user equipment (UE) gateway moves into the coverage area of the sensor nodes, broadcasts beacon packets to these nodes, and provides backhaul access to these WSN nodes. MCN can send the detected WSN data directly to a central data center~\cite{15}. The main disadvantages of WSNs include less mobility robustness, small coverage, and weak terminals. MCNs, on the other hand, provide more layer control, a longer network lifetime, and quality of service (QoS) for WSNs applications and have the advantages of reliable mobility, broad coverage, and powerful user terminals, but are expensive and difficult to deploy and manage~\cite{16}. The convergence of MCN and WSN can benefit each other: MCNs provide a greater level of layer control and optimization to increase network life and WSN performance and to provide better QoS with the use of WSNs; WSNs may operate as cognitive and intelligent enablers of MCNs, wireless services and increasingly data-centric applications are enabled by the architecture of WSN and MCN convergence networks, MCNs have the potential to make WSNs more energy efficient and improve network performance, and mobile MCN terminals act both as sensor nodes and gateways for WSNs~\cite{17}. In addition, sensor nodes in WSN and MCN convergence networks collect data and send it to a data server through MCN~\cite{18}.
\subsection{System Architecture Design}
\begin{figure}[!h]
		\centering 	\includegraphics[scale=0.35]{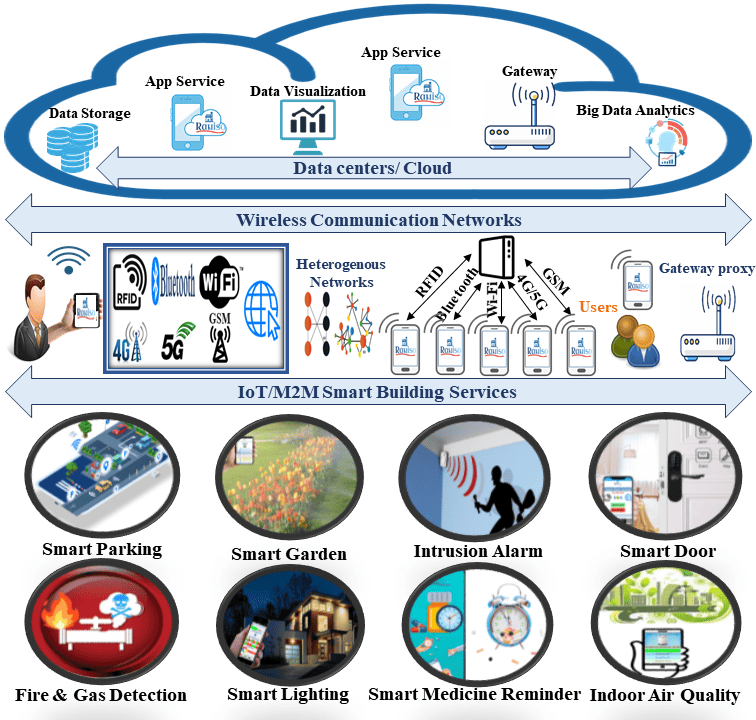}
		\caption{IoT/M2M smart building architecture.}
		\label{fig1}
	\end{figure}

The proposed architecture design for IoT/M2M smart building system based on heterogeneous networks uses different known hardware to collect data and manage it according to a business’ functions and services. The main services offered are smart parking, garden irrigation automation, intrusion alarm, smart door, fire and gas detection, smart lighting, smart medication reminder, and indoor air quality monitoring. All these services are designed and implemented to remotely control and monitor the building through the Raniso App via RFID/Bluetooth/Wifi connectivity and cellular networks such as GSM, 4G, or 5G. This IoT/M2M smart building infrastructure design helps owners, operators, and facility managers improve asset reliability and performance and is beneficial in preventing the loss of resources and human lives caused by undesired events. Moreover, the proposed system is energy-efficient and low-cost that can be used in various buildings, including hospitals, hotels, universities, businesses, etc. Figure~\ref{fig1} shows the proposed architectural design.

\subsection{System Hardware and Software}
In the proposed IoT/M2M smart building system based on {HetNets} networks, we use different sensors and actuators to collect data and manage it according to services, where the Arduino microcontroller represents the brain. Besides, the NodeMCU board and other modules like Bluetooth HC-06, Sim800l, and RFID are used for wireless communication. The main hardware components used are described in Table~\ref{Table1} whereas the software used is defined in Table~\ref{Table2}. Next, we discuss the design and implementation of the proposed system.

\begin{table*}[!htb]
\caption{\MakeUppercase{The main specifications of hardware used.}}
\centering
\resizebox{2\columnwidth}{!}{%
\begin{tabular}{|p{1.8cm}|p{2.5cm}|p{17cm}|}
\hline
\textbf{Components} &
 \textbf{Type} &
 \textbf{Specifications} \\ \hline
 {Arduino} &
 Mega &
 It is an ATmega2560-based microcontroller board. There are 54 digital input/output pins, 16 analog inputs, 4 UARTs (hardware serial ports), a 16 MHz crystal oscillator, a USB connection, a power jack, an ICSP header, and a reset button on this board \cite{19}.\\ \hline
 {Arduino} &
 UNO &
 It is an open-source microcontroller board based on the ATmega328 processor and it is created by “Arduino.cc”. The board contains 14 digital input/output pins, six analog inputs, a 16 MHz ceramic resonator, a USB connection, a power jack, an ICSP header, and a reset button \cite{20}.\\ \hline
 {NodeMCU} &
 V3 &
 It is an open-source board comprised of a physical programmable circuit board \cite{21} created by a Chinese firm named Espressif makes a System on Chip (SoC) called ESP8266. This board is defined as a low-cost Wi-Fi chip in which uses the Lua programming language and includes a full TCP/IP stack and a microcontroller.\\ \hline
 {Modules} &
 RC522 RFID & The term RFID stands for Radio Frequency Identification. The RFID reader can be used to read and retrieve information from RFID cards.
\\ \hline
{Modules} &
 SIM800L & It is a miniature GSM modem that helps in making calls and sending messages and it can access the internet through GPRS, TCP/IP. This module can be used in a wide variety of IoT/M2M projects.
\\ \hline
{Modules} &
 GPS  & The global positioning system (GPS) module uses satellite technology to determine continuously data like longitude, latitude, speed, distance traveled, etc.
\\ \hline
{Modules} &
 SD Card  & It enables communication with the memory cards as well as the writing and reading of data on them.
\\ \hline
{Modules} &
 Bluetooth HC-06 & is a basic Bluetooth SPP (Serial Port Protocol) module that is used to transmit data within a small area in the ISM band between 2.4 and 2.485 GHz \cite{26}.
\\ \hline
{Modules} &
 DS3231 & The DS3231 Real Time Clock (RTC) is a low-power clock/calendar with a battery backup SRAM of 56 bytes. The clock/calendar gives certified data for seconds, minutes, hours, days, dates, months, and years.
\\ \hline
{Sensors} &
 TCRT5000 & It is a sensor that detects infrared light. It is simply a photodiode and a phototransistor combined \cite{7, Wei2022}.
\\ \hline
{Sensors} &
 Ultrason HC-SR04 & It is an electronic device that calculates the distance between physical objects using sonar. It has a wide object detecting range of 2 cm to 400 cm with great accuracy \cite{9}.
\\ \hline
{Sensors} &
 Soil moisture & It measures the content of water and humidity in the soil. As the soil moisture increases, the sensor allows current to pass through, and the greater the moisture, the greater the proportion of the connection \cite{26, Wenjun2022}.
\\ \hline
{Sensors} &
 DHT11 & It is a digital sensor that detects the temperature and humidity of the place. It provides high levels of dependability, accuracy, and long-term stability \cite{7}.
\\ \hline
{Sensors} &
 SW-420 Vibration  & It measures the vibration level and tilts motion continuously. This sensor is used for security purposes.
\\ \hline
{Sensors} &
 MQ2 & It is a gas detection sensor that senses or measures various types of gasses like LPG, Alcohol, Propane, H2, CO, and even methane \cite{26}.
\\ \hline
{Sensors} &
 Flame & It is designed to find out if there is a flame as well as regular light with a detection range of up to 100 cm and a wavelength ranging from 760 to 1100 nm.
\\ \hline
{Sensors} &
 PIR HC-SR501 & PIR sensor refers to Passive Infrared that is designed to sense the motion reliant on the infrared generated by the human body and other living creatures.
\\ \hline
{Sensors} &
 MQ135 & It is a gas sensor used for air quality in which it detects or measures NOx, Nh3, CO2, Benzene, Alcohol, and Smoke.
\\ \hline
{Actuators} &
 Servo motor & It is an electrical engine that can be utilized to push or rotate an object with high precision [15]
\\ \hline
 {Actuators} &
 LCD  & It is a type of tool for displaying characters in which the primary viewer is a liquid crystal [5].
\\ \hline
{Actuators} &
 Fan & It is a device used to create air movement for ventilation in hot climates to create a relaxing environment.
\\ \hline
{Actuators} &
 Keyboard & It is a contact matrix of buttons made up of rows and columns. It is used to enter numbers and letters.
\\ \hline
{Actuators} &
 Speaker& It consists of power amplification and voice outputs, it is used to emit a variety of sounds.
\\ \hline
{Actuators} &
 Buzzer & It is an audio signaling gadget that produces sound. In the proposed system, the buzzer is utilized as a notification alarm.
\\ \hline
{Actuators} &
 LDR   & Light Dependent Resistor (LDR) is an electrical component whose resistance varies in response to the perceived light \cite{7}.
\\ \hline
{Actuators} &
 LED   & A light-emitting diode (LED) is a semiconductor light source that emits erratic radiation as a result of spontaneous photon emissions. It is utilized to represent light \cite{9}.
\\ \hline
\end{tabular}%
}
\label{Table1}
\end{table*}

\begin{table*}[!htb]
\caption{\MakeUppercase{The main specifications of software used.}}
\centering
\resizebox{2\columnwidth}{!}{%
\begin{tabular}{|p{2.5cm}|p{19cm}|}
\hline
\textbf{Software} &
 \textbf{Specifications} \\ \hline
 {Arduino IDE} &
 It is an open-source software written in Java and based on the processing programming language. It allows to write, modify and convert programs directly to microcontrollers \cite{9}. The version used in this work is version 1.8.16.\\ \hline
 
 {Proteus} & It is an open-source software for virtual system modeling and circuit simulation. It is mainly used for electronic design automation. The version used is version 8.13. 
 \\ \hline
 
 {Raniso App} & It is a mobile application first we created in 2019. It acts as an intermediary between the smartphone and many hardware components and it is used to control several electrical devices via various wireless networks. In this research work we use ten interfaces of Raniso App (see Figure~\ref{fig2}.) to control and monitor the services, where one interface has been created in this app for each proposed service of IoT/M2M smart building.
 \\ \hline
\end{tabular}%
}
\label{Table2}
\end{table*}
 
	\begin{figure}[!h]
		\centering 
		\includegraphics[scale=0.35]{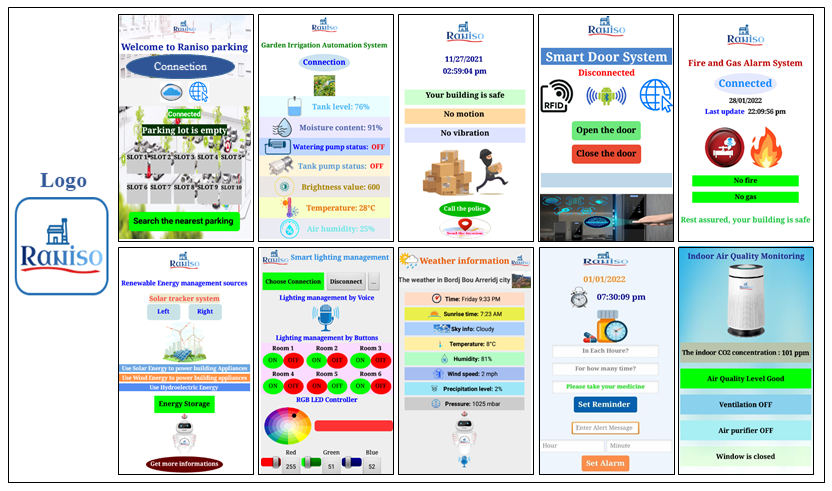}
		\caption{The different interfaces of the Raniso App used for the proposed IoT/M2M smart building.}
		\label{fig2}
	\end{figure}

\section{Designing and Implementing IoT/M2M Smart Building services}
This section is devoted to the simulations and practical implementation of IoT/M2M smart building services. There are two different forms of main processing units used in the deployment of all these services, which are Arduino and NodeMCU. The Arduino will be used as the controller of the devices, while NodeMCU will be the controller of the phone devices so that it can control the appliances from far away. In the following, we discuss each smart application prototype explicitly.
 
 \subsection{Smart Parking}
\begin{figure}[!h]
		\centering 
		\includegraphics[scale=0.3]{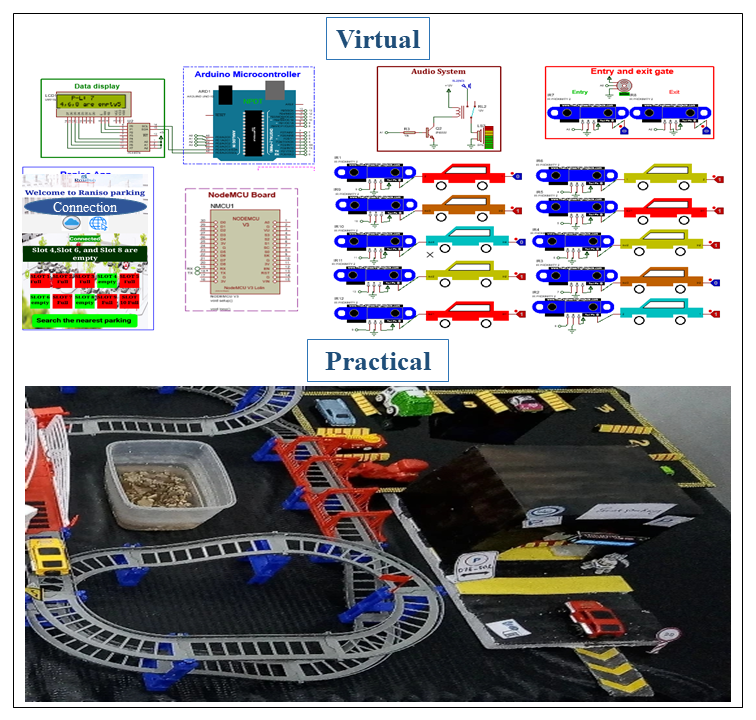}
		\caption{The proposed system prototype for smart parking.}
		\label{fig3}
	\end{figure}

The smart parking involves an IoT/M2M-based system that sends data on the availability of all parking places in real-time and selects the optimal one. The system comprises an Arduino Uno microcontroller, NodeMCU board, infrared sensors (TCRT5000), servo motor, LCD, speaker, and a battery. Figure~\ref{fig3} shows the proposed system prototype for smart parking.

Using infrared sensors, the system identifies whether parking spaces are occupied, then scans the number of available spaces and updates data with the cloud server every 30 seconds. The parking slot availability may be checked online from anywhere, and perform hassle-free parking through the smart parking interface of the Raniso App. Also, the LCD placed in front of the gate will show which slot is free; besides, when a vehicle is detected in a specific slot, the corresponding LED in the Raniso App lights up. If all spaces are occupied by vehicles, the parking gate will not open, and the audio system will say: "sorry, this parking lot is full, you can search for another available parking lot via the Raniso app." If any spaces are available, the parking gate will open, allowing the vehicle to pass, and the audio system will welcome the user. As a result, this system solves the city parking problem and provides users with a reliable M2M/IoT-based parking management solution. With this solution, users can easily find available, nearby and cheap parking using the interface of the Raniso app equipped with GPS technology, wherever they are and whenever they want as shown on Figure~\ref{fig4}. The following Figure shows the overall proposed solution for a smart parking system.
\begin{figure}[!h]
		\centering 
		\includegraphics[scale=0.145]{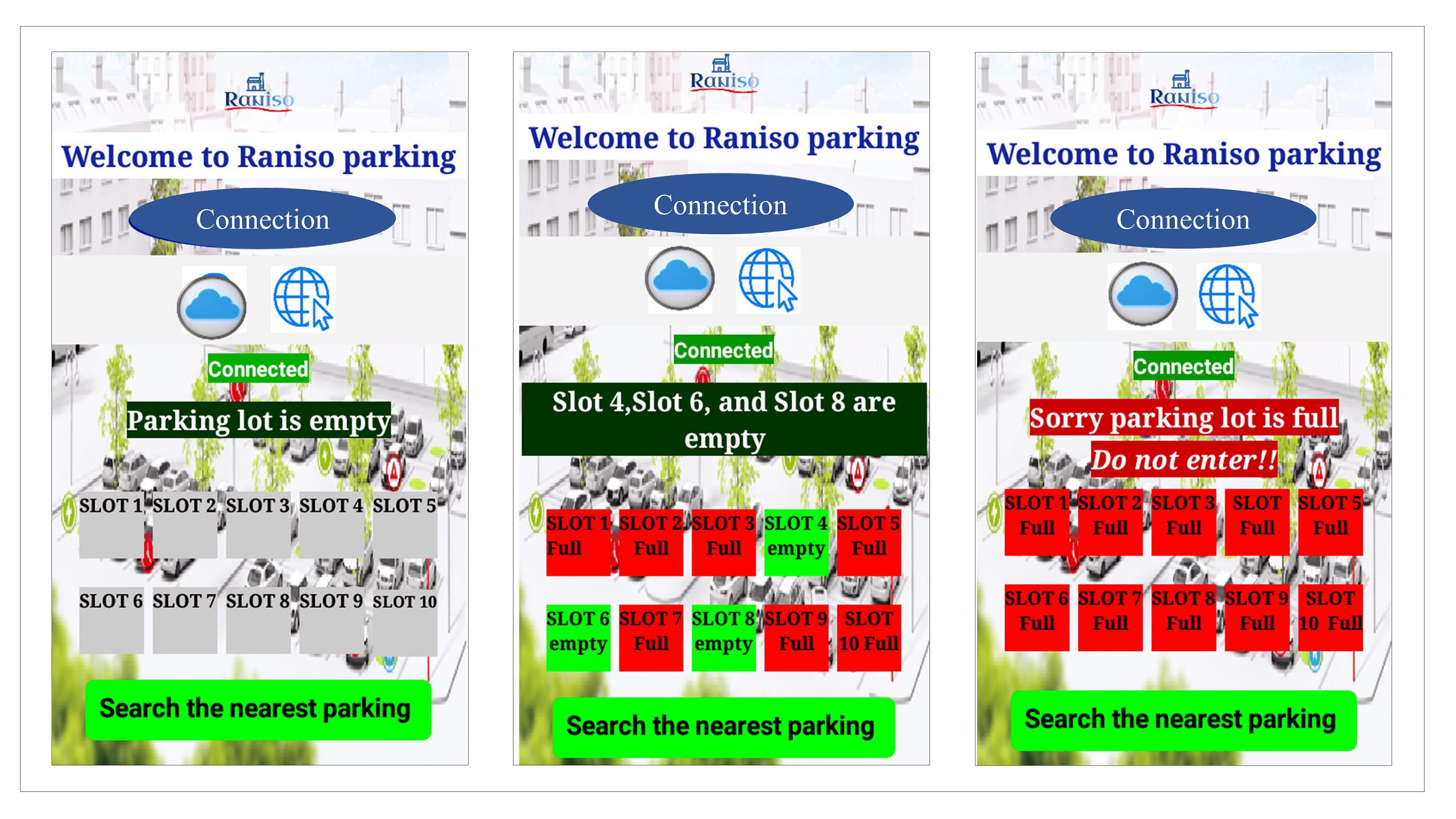}
		\caption{Smart parking interface on the Raniso App.}
		\label{fig4}
	\end{figure}
	
	\begin{figure}[!h]
		\centering 
		\includegraphics[scale=0.1475]{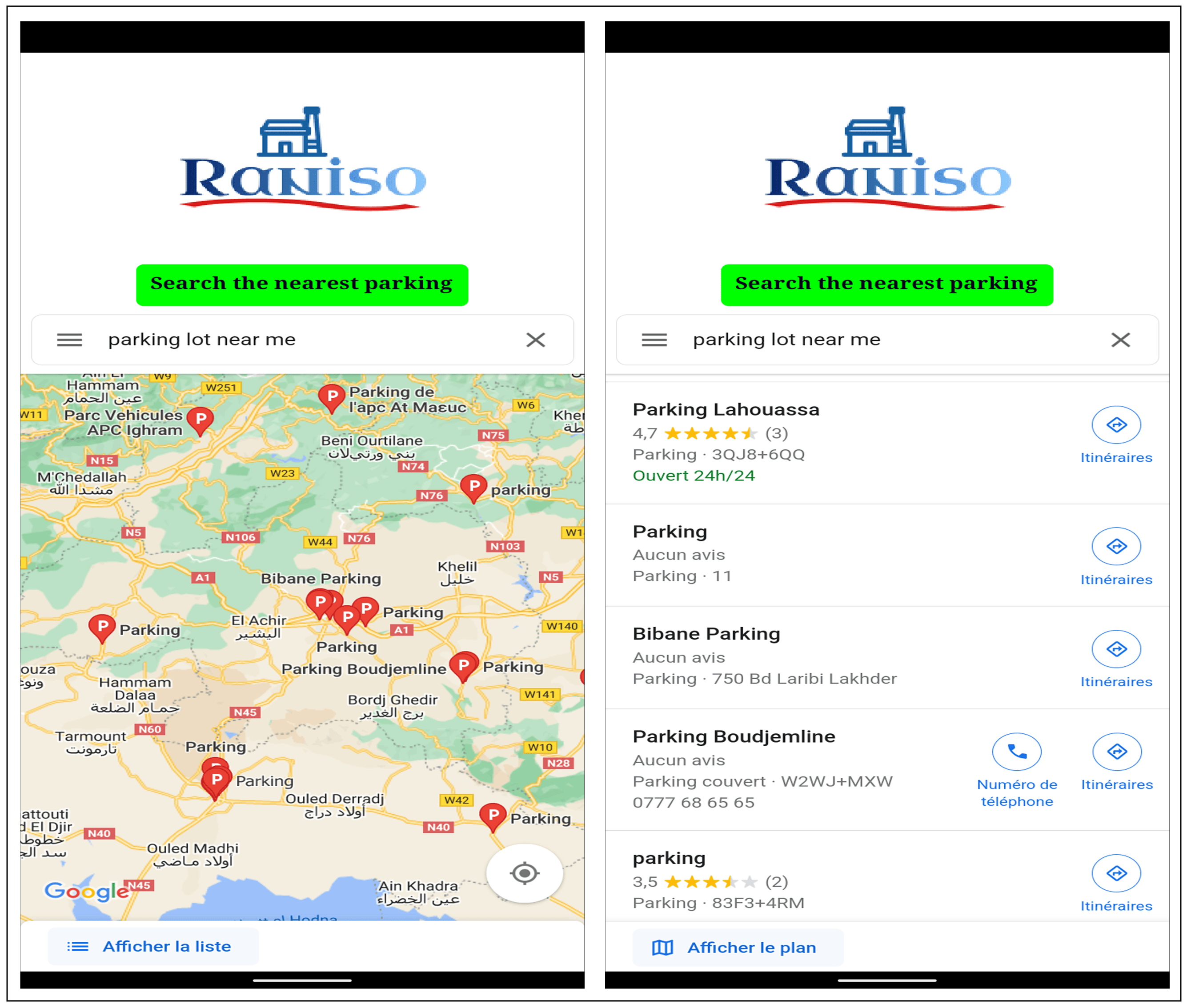}
		\caption{Search results for car parks near the user's location on Raniso App.}
		\label{fig5}
	\end{figure}

\subsection{Garden Irrigation Automation System}
\begin{figure}[!h]
		\centering 
		\includegraphics[scale=0.375]{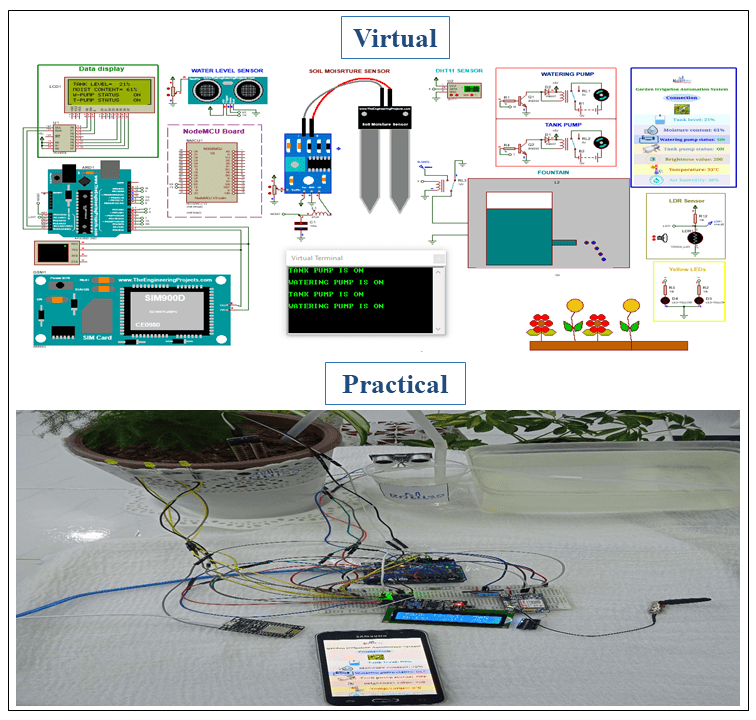}
		\caption{Illustration of the garden irrigation automation system.}
		\label{fig6}
	\end{figure}

\begin{figure}[!h]
		\centering 
		\includegraphics[scale=0.15]{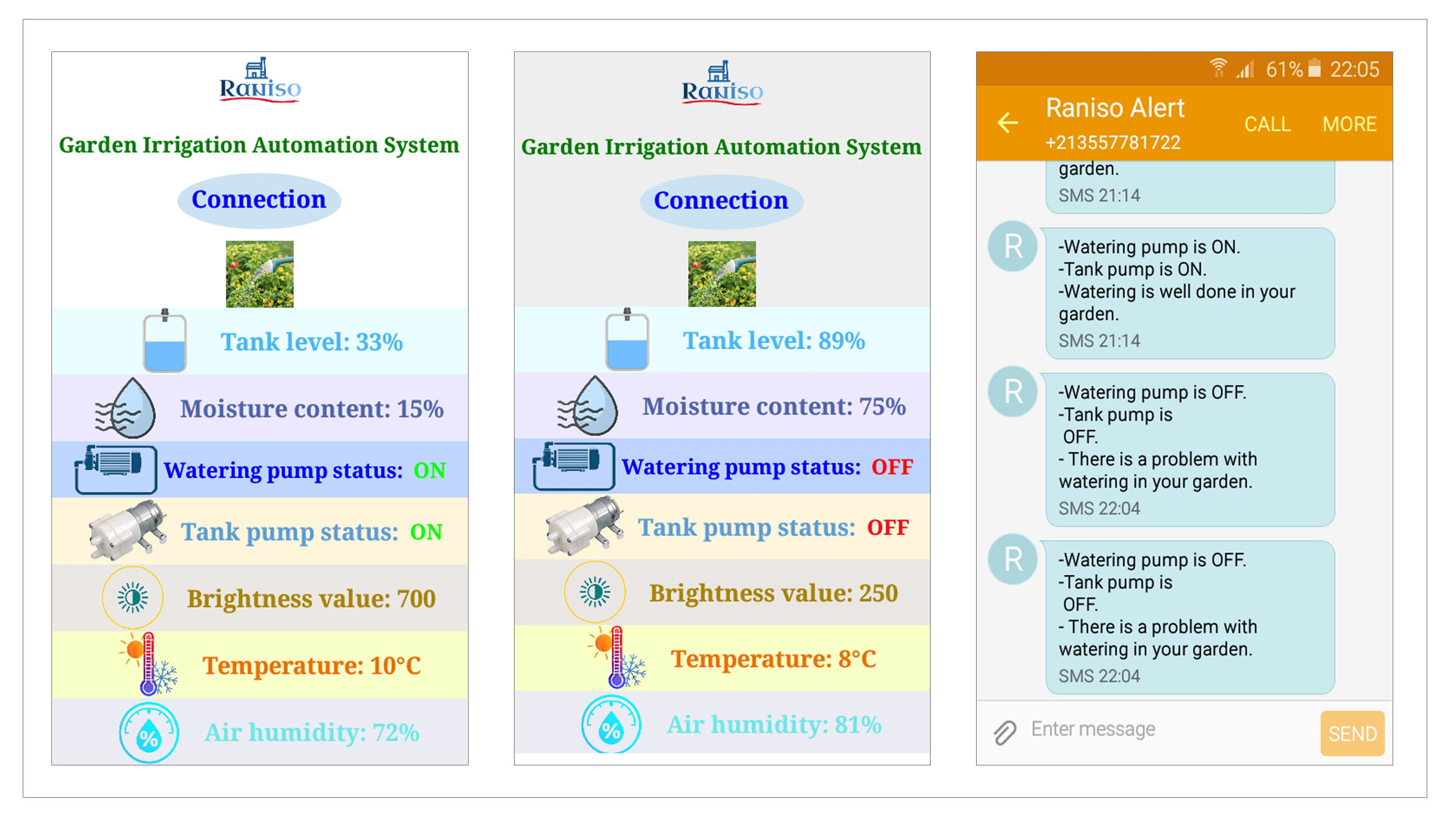}
		\caption{Garden irrigation automation interface on the Raniso App.}
		\label{fig7}
	\end{figure}

\begin{figure}[!h]
		\centering 
		\includegraphics[scale=0.31]{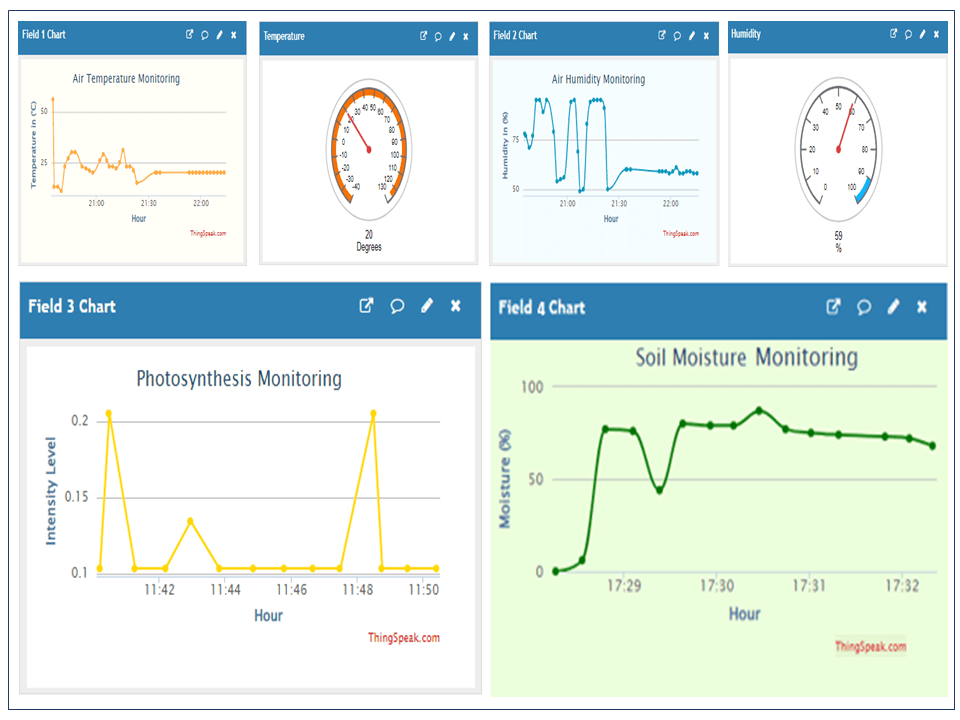}
		\caption{Garden Irrigation Monitoring System on ThingSpeak.}
		\label{fig8}
	\end{figure}

The proposed irrigation automation system for gardens refers to the system's functionality with no or just a minimum of manual intervention, where it monitors the plant's temperature, humidity, light levels, and soil moisture (see Figure~\ref{fig6}). The system includes a soil moisture sensor which is used to measure the percentage of soil moisture to optimize the irrigation dosage to eliminate water waste, a DHT11 sensor is used to monitor temperature and humidity amounts, an LDR sensor to sense the intensity of light, two yellow LEDs for controlling the photosynthesis process during the night to make the plant grow faster, the ultrasonic sensor (HC-SR04) to measure the level of water inside the tank, GSM module to make calls and send text messages to the farmer when the pump changes its state or when there is any watering problem, the LCD screen is used to display the level of water and moisture content together with the pump status, and two pumps one for watering the plants and the second one for supplying water to the tank. If the soil is dry, the irrigation fountain operates, watering the garden at the time of need and switches OFF when the soil is wet to save water. 
{In addition, this solution allows for remote control and monitoring of the garden irrigation system via the Raniso app (see Figure~}\ref{fig7}),{ and via our channel in ThingSpeak which is used to view data from the proposed smart garden system remotely (see Figure}~\ref{fig8}),{ making it easier to manage the irrigation systems and make necessary adjustments in real-time. }
This system optimizes resources (water, energy, and fertilizers) and irrigation scheduling through an intelligent monitoring system.

\subsection{Intrusion Alarm System}
\begin{figure}[!h]
		\centering 
		\includegraphics[scale=0.375]{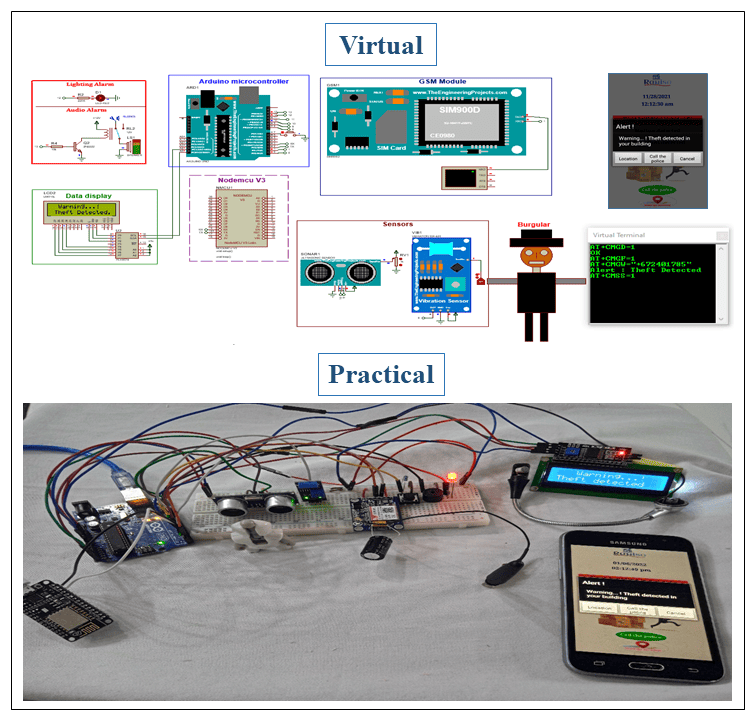}
		\caption{Intrusion alarm system.}
		\label{fig9_1}
	\end{figure}

\begin{figure}[!h]
		\centering 
		\includegraphics[scale=0.175]{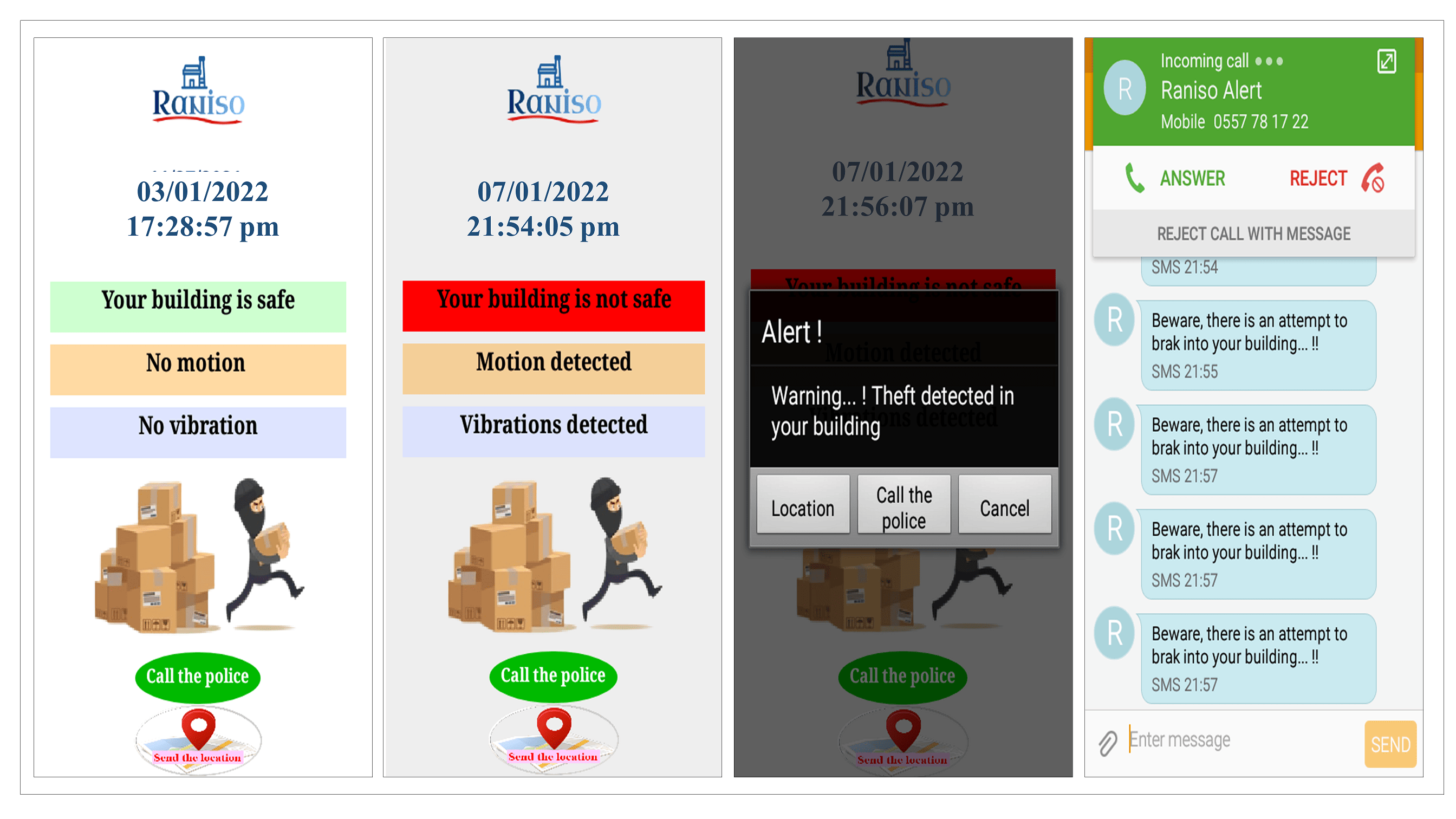}
		\caption{Intrusion alarm interface on the Raniso App.}
		\label{fig10_1}
	\end{figure}

\begin{figure}[!h]
		\centering 
		\includegraphics[scale=0.06]{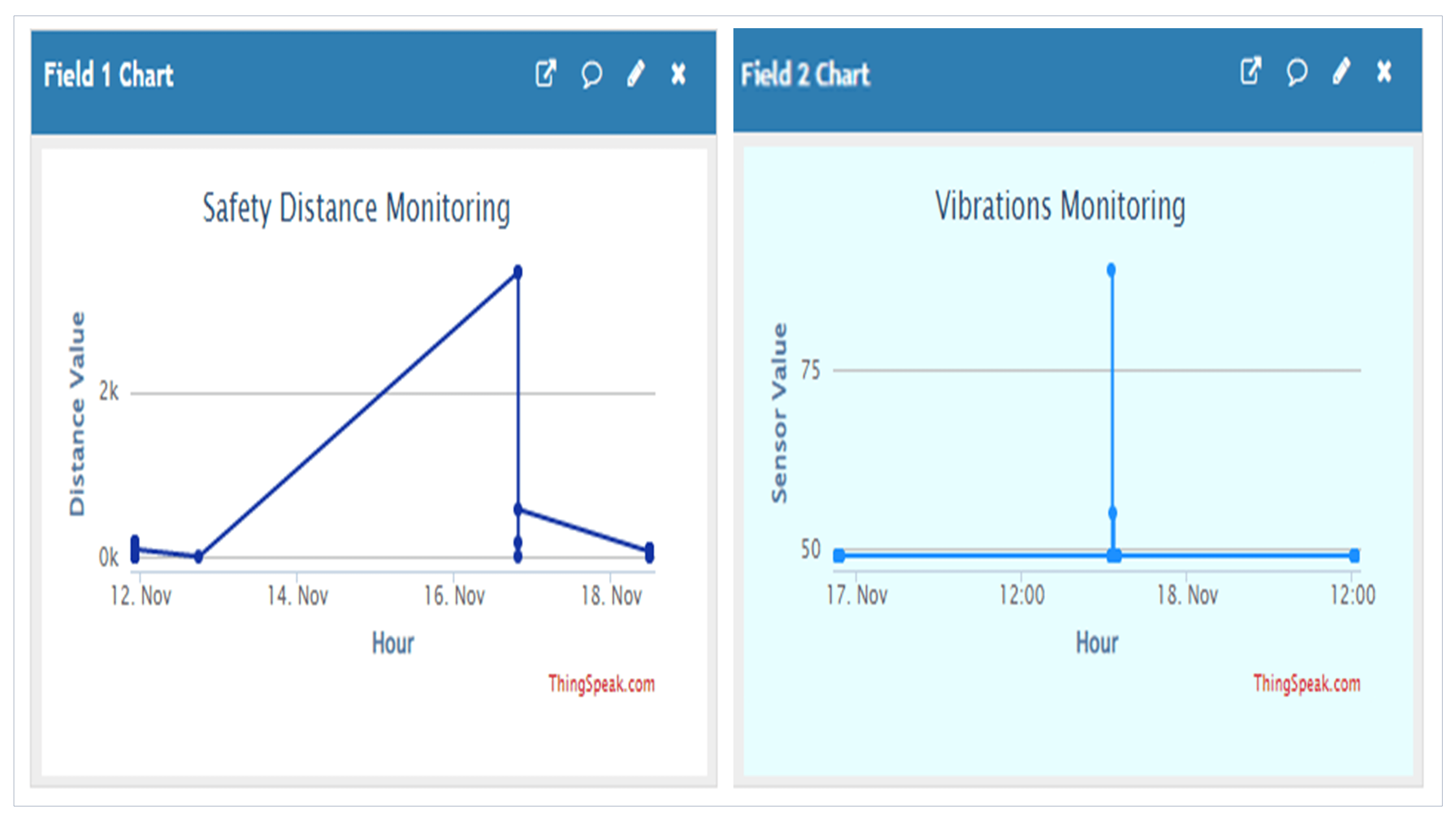}
		\caption{Intrusion alarm system monitoring on Thingspeak.}
		\label{fig11_1}
\end{figure}

The increased number of robberies in buildings makes people worried about losing their property. Hence, the proposed service presented in  Figure~\ref{fig9_1} enables the protection of the building from theft or attack by using a GSM module, ultrasonic sensor, vibration sensor, speaker, and red LED. The ultrasonic sensor is used to capture the distance and the vibration sensor to measure the vibrations. When the distance is close and vibrations are detected, the GSM module sends calls and short messages to the owner of the building, the LCD screen shows an alarm state, the red LED lights up, and the speaker emits a sound indicating that there is a theft, where the resulting sound is obtained from a micro SD memory card in MP3 format, which allows users to use the sound of screams, a guard dog or a siren to scare away the thief. {Additionally, users can monitor their building from anywhere at any time via our channel in the ThingSpeak platform as illustrated in } Figure~\ref{fig11_1}, and via the intrusion alarm interface in the Raniso app
that sends real-time alerts in case of any danger as shown in Figure~\ref{fig10_1}.

\subsection{Smart Door System}
\begin{figure}[!h]
		\centering 
		\includegraphics[scale=0.35]{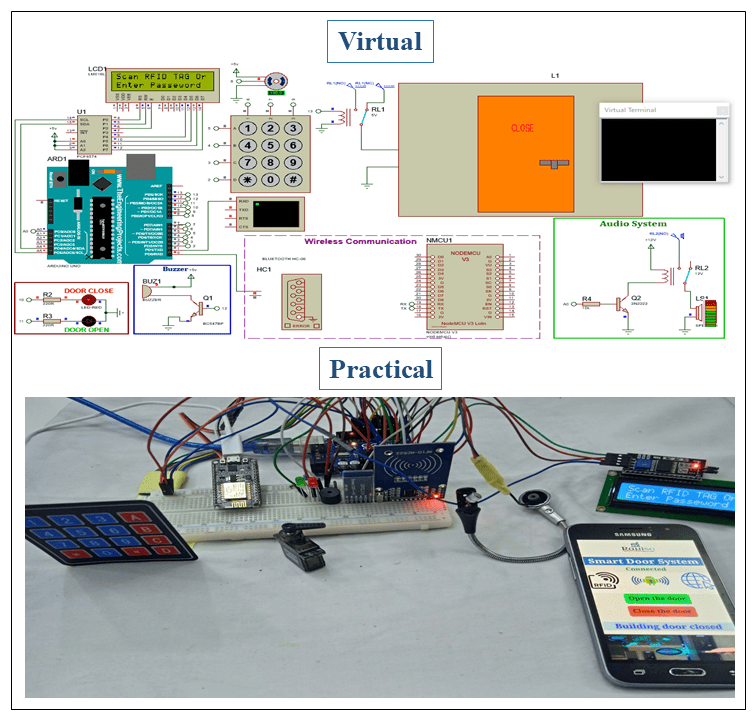}
		\caption{Prototype of a smart door system.}
		\label{fig12_1}
	\end{figure}

\begin{figure}[!h]
		\centering 
		\includegraphics[scale=0.0625]{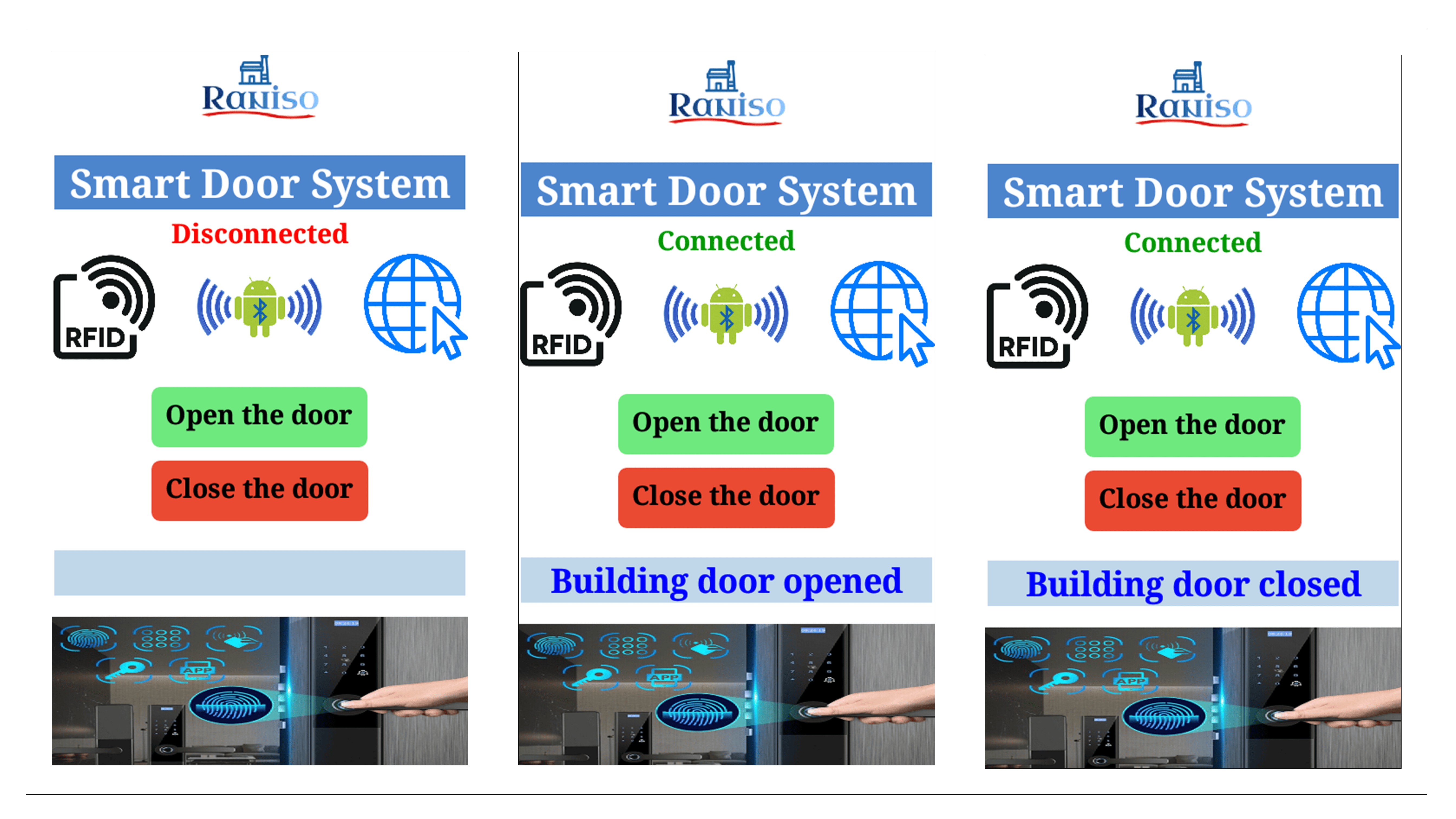}
		\caption{Smart door interface on the Raniso App.}
		\label{fig13_1}
	\end{figure}

The proposed smart door system is an automatic identification and authentication system deployed at the doors of various buildings, including banks, corporate offices, financial institutions, jewelry stores, government organizations, etc. It is designed to prevent unauthorized access and violation by using a Bluetooth module, RFID module, servo motor, keypad, LCD screen, speaker, buzzer, and LEDs. As soon as the user approaches the door, the sound system reminds him to disinfect his hands and take all precautions against COVID-19. The LCD screen also displays all the options by which the user can control the keyless door, either by RFID card, password, or remotely via the smart door system interface of the Raniso app using the Bluetooth connection or the Internet via WiFi, 4G/5G (see Figure~\ref{fig13_1}). The system will give access using the Raniso app, or on scanning the right tag or entering the correct password, and on scanning the wrong tag or entering the wrong password, the system will deny, a red LED will light up, and the buzzer will make a beep sound. This solution is intended to control doors in the building with a relatively low-cost design, user-friendly interface, and ease of installation. It also provides a protection system and building security. Figure~\ref{fig12_1} shows the prototype of the proposed smart door system.

\subsection{Fire and gas alarm system}
\begin{figure}[!h]
		\centering 
		\includegraphics[scale=0.425]{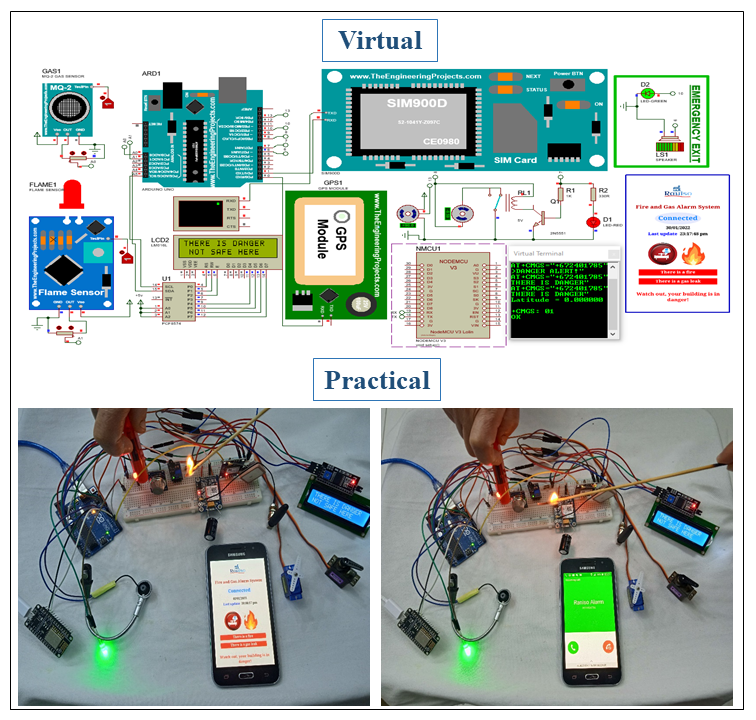}
		\caption{Fire and gas alarm system.}
		\label{fig14_1}
	\end{figure}

\begin{figure}[!h]
		\centering 
		\includegraphics[scale=0.625]{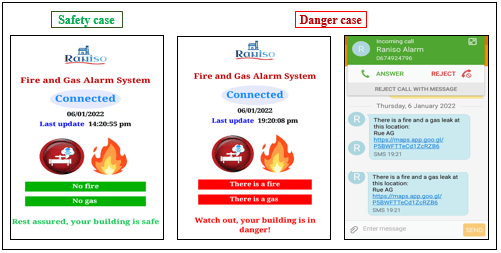}
		\caption{Fire and gas detection interface on the Raniso App.}
		\label{fig15_1}
	\end{figure}

\begin{figure}[!h]
		\centering 
		\includegraphics[scale=0.15]{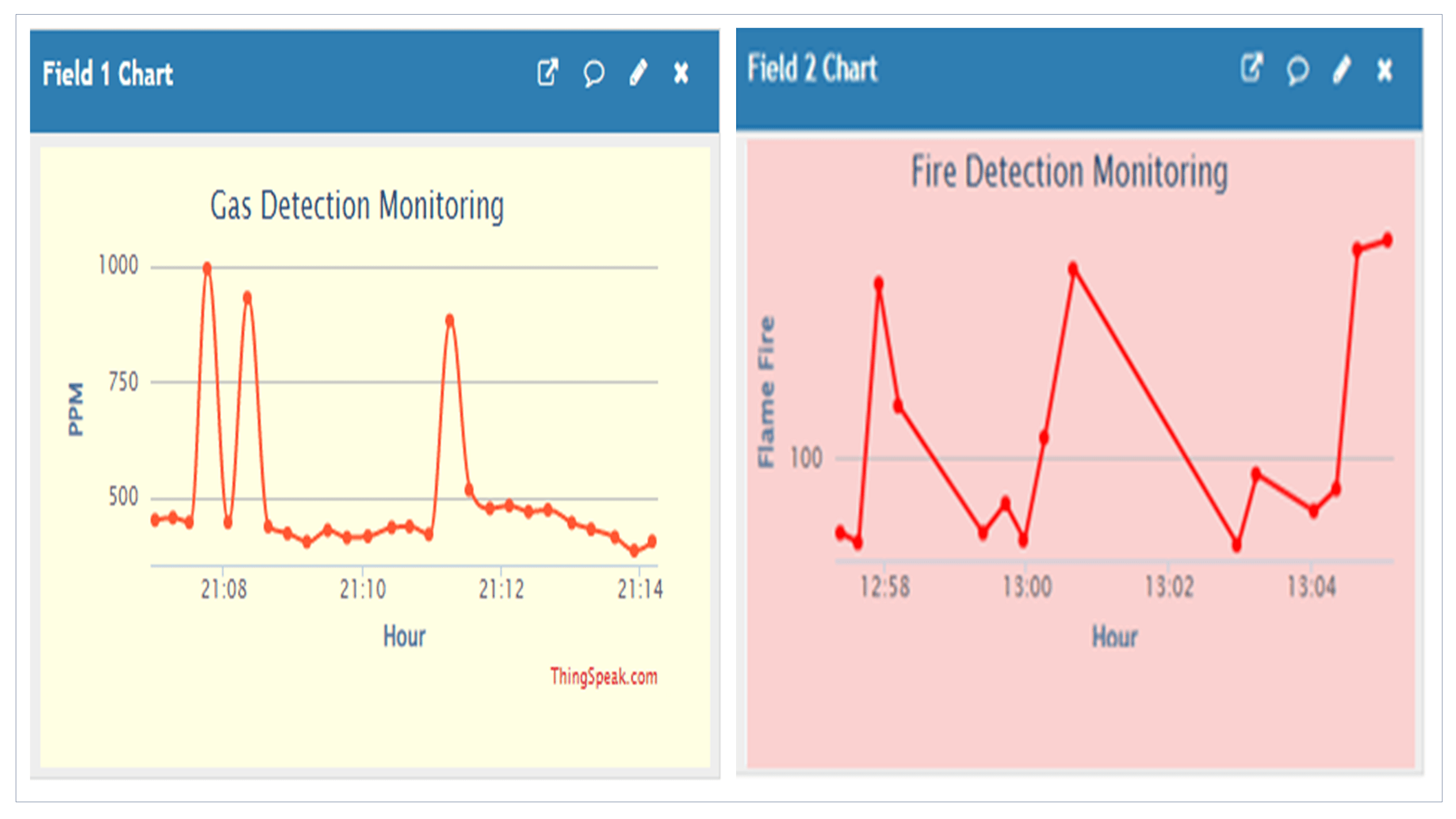}
		\caption{Fire and gas detection monitoring on ThingSpeak.}
		\label{fig16_1}
	\end{figure}

This proposed system for extinguishing fires and detecting gas leaks is implemented to protect lives and property from gas and fire hazards that lead to serious accidents, human injuries, and material losses (see Figure~\ref{fig14_1}). This service is based on a GSM module, GPS shield, servo motors, a flame sensor, Gas leakage sensor, LCD screen, speaker, and LEDs. Sensors are used to sense fire and gas in the building. Whenever the fire or the gas is detected, the alarm rings, the LCD screen displays ”There is Danger, Not safe here”, a red LED lights up, one servo motor opens the water sprayer and the other one opens the window. In addition, the system not only sends calls and SMS alerts to the owner but also sends an SMS along with the location of the incident to the fire station and civil protection. The system also guides the building owner to a safe, fire-free route through an audio system and a green light. {Besides, users can monitor data of gas and fire detection of their buildings from anywhere at any time via our channel in the ThingSpeak platform as shown in Figure}~\ref{fig16_1}, {and via the Raniso app’s fire and gas alarm interface which sends real-time alerts in case of fire and gas leaks as illustrated in }Figure~\ref{fig15_1}.

\subsection{Smart Lighting Management }
\begin{figure}[!h]
		\centering 
		\includegraphics[scale=0.4]{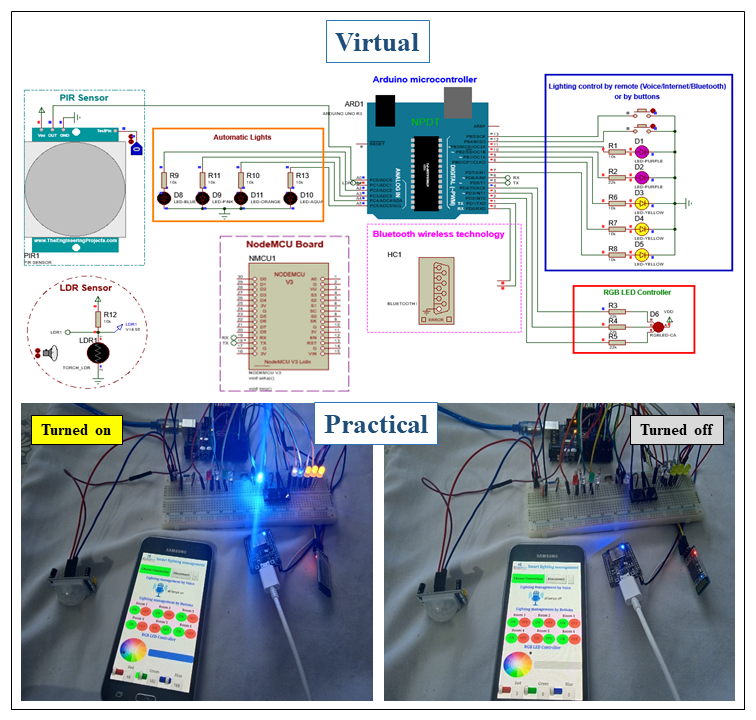}
		\caption{Smart lighting management by Remote (Voice/Internet/Bluetooth) or by Push Button Switches.}
		\label{fig17_1}
	\end{figure}

The proposed system is advanced smart lighting management based on IoT/M2M technologies and was built using a Bluetooth module, PIR sensor, LEDs, RGB LEDs, LDR sensors, and push-buttons as shown in Figure~\ref{fig17_1}. 
	
\begin{figure}[!h]
		\centering 
		\includegraphics[scale=0.4]{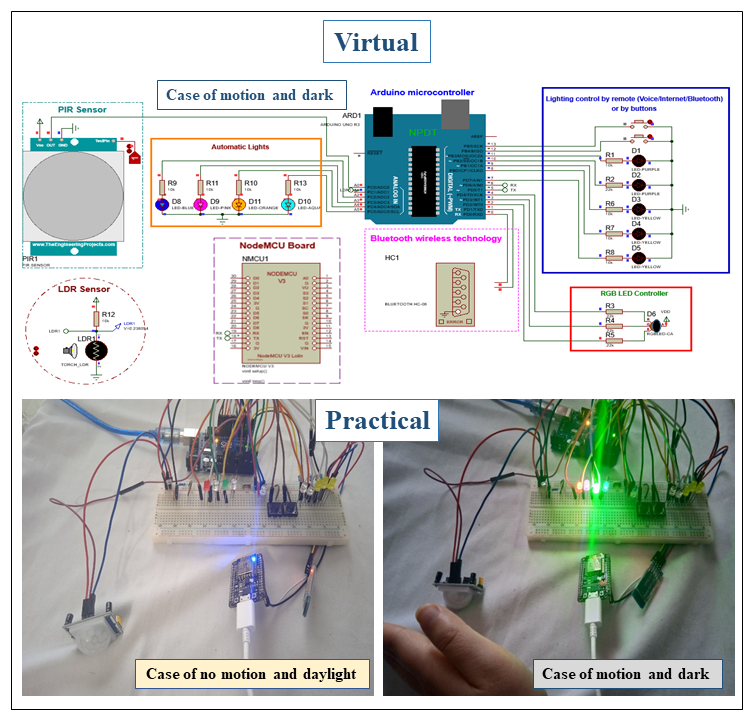}
		\caption{Automatic lights.}
		\label{fig18_1}
	\end{figure}

\begin{figure}[!h]
		\centering 
		\includegraphics[scale=0.15]{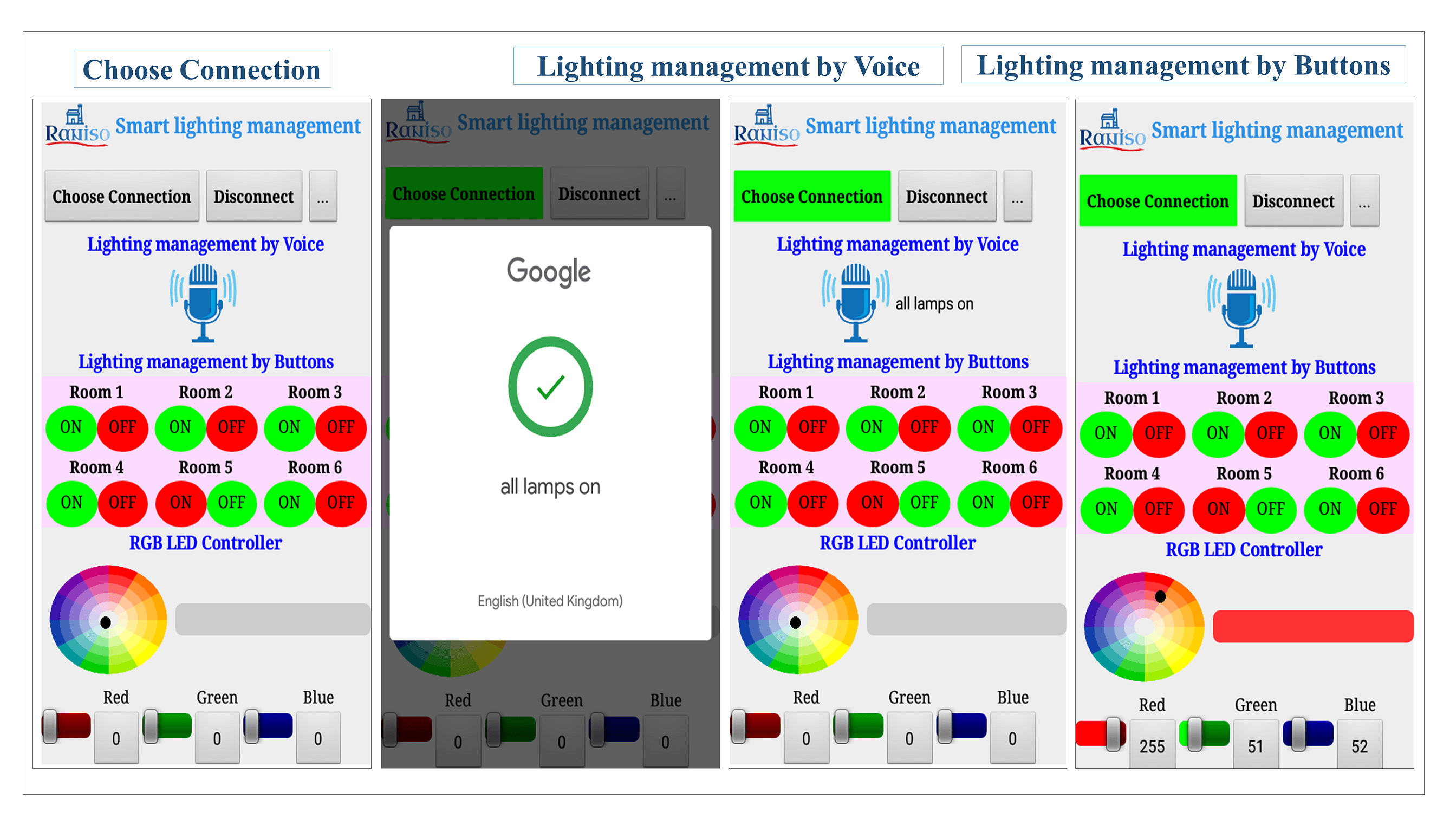}
		\caption{Smart lighting interface on the Raniso App.}
		\label{fig19_1}
	\end{figure}

\begin{figure}[!h]
		\centering 
		\includegraphics[scale=0.06]{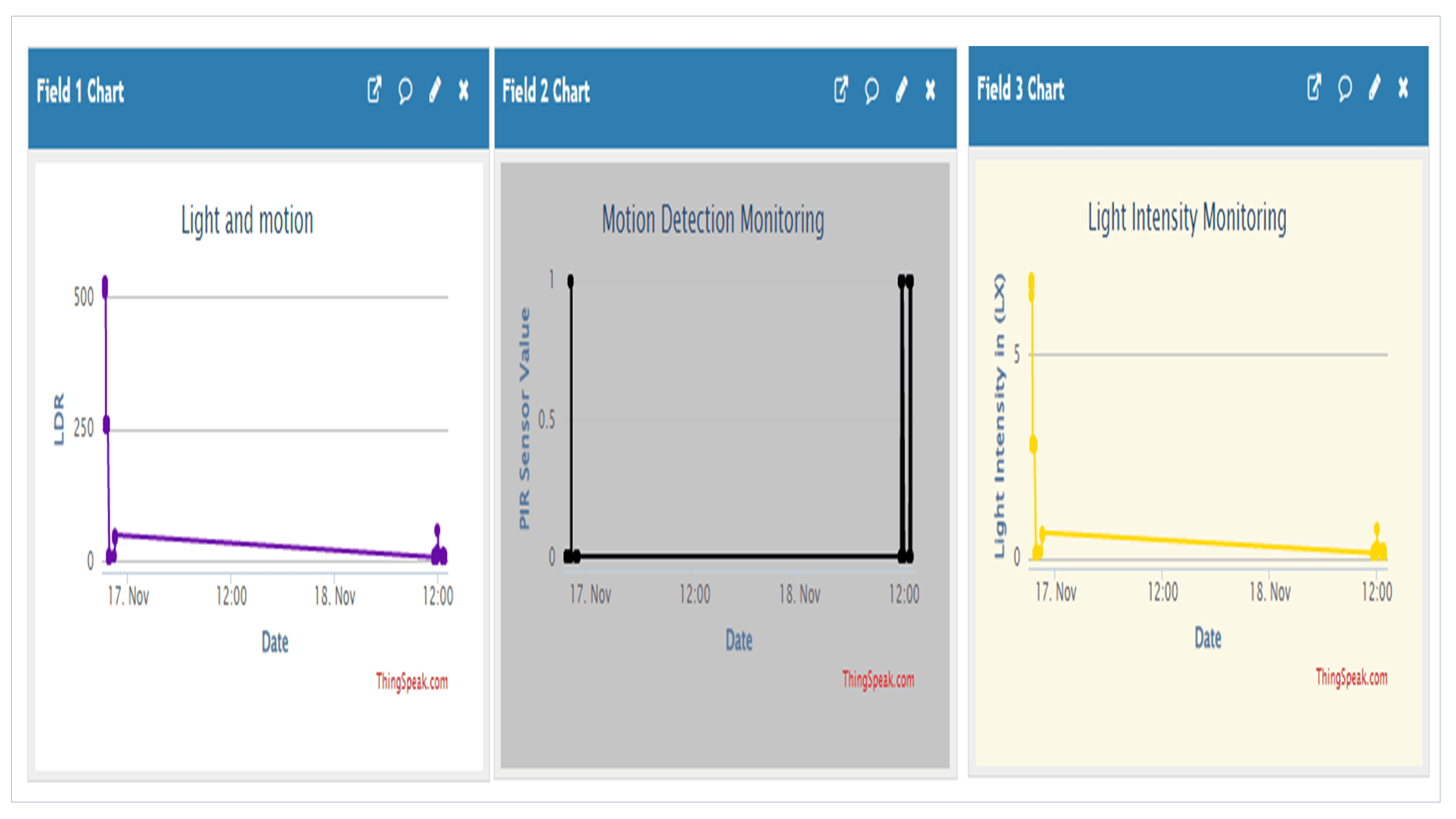}
		\caption{Smart lighting monitoring on ThingSpeak.}
		\label{fig20_1}
	\end{figure}

This system is built to be energy-efficient, convenient, and safe. It gives users the freedom to choose how to control the lights, whether controlling the lights via pushbuttons, Bluetooth, Wi-Fi, 4G/5G networks, or via the use of artificial intelligence by enabling Google Assistant voice commands integrated into the Raniso App. It also suggests to the user the best way to control the lights in real-time. Also, we can generate any color with an RGB LED by adjusting the brightness of the individual Red, Green, and Blue LEDs through the interface of smart lighting in the Raniso App as shown in Figure~\ref{fig19_1}. Besides, in such places as garages, stairs, bathrooms, etc. where there is no need for continuous light, we have implemented automatic lights illustrated in Figure~\ref{fig18_1}. Where lights are automatically turned ON when a human is present and when it is dark. { Additionally, the user can monitor in real-time LDR sensor data, motion detection, and light intensity via our channel in ThingSpeak as shown in Figure}~\ref{fig20_1}. This service was proposed to save the max of energy and provide a level of comfort and convenience. The following figure represents the proposed smart lighting management by Remote (Voice/Internet/Bluetooth) or by Push Button Switches.

\subsection{Smart Medicine Reminder}
\begin{figure}[!h]
		\centering 
		\includegraphics[scale=0.4]{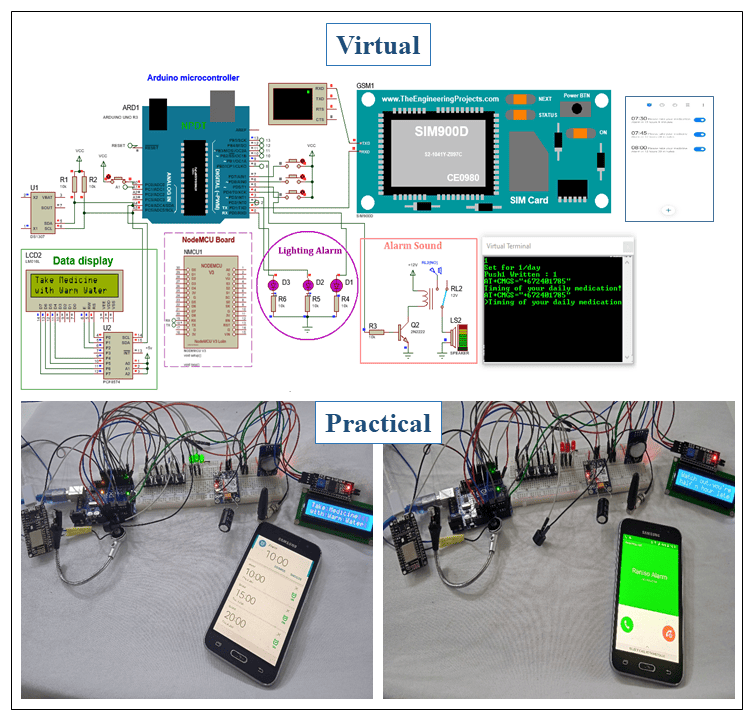}
		\caption{Smart medicine reminder.}
		\label{fig21_1}
	\end{figure}

Medicines help us live longer and healthier. But taking it the wrong way, mixing certain medications, or forgetting to take it at the right time can be dangerous. The proposed IoT/M2M Smart Medicine Reminder solves such problems by reminding and alerting patients to take the proper dose at the right moment (see Figure~\ref{fig21_1}).
	
\begin{figure}[!h]
		\centering 
		\includegraphics[scale=0.06]{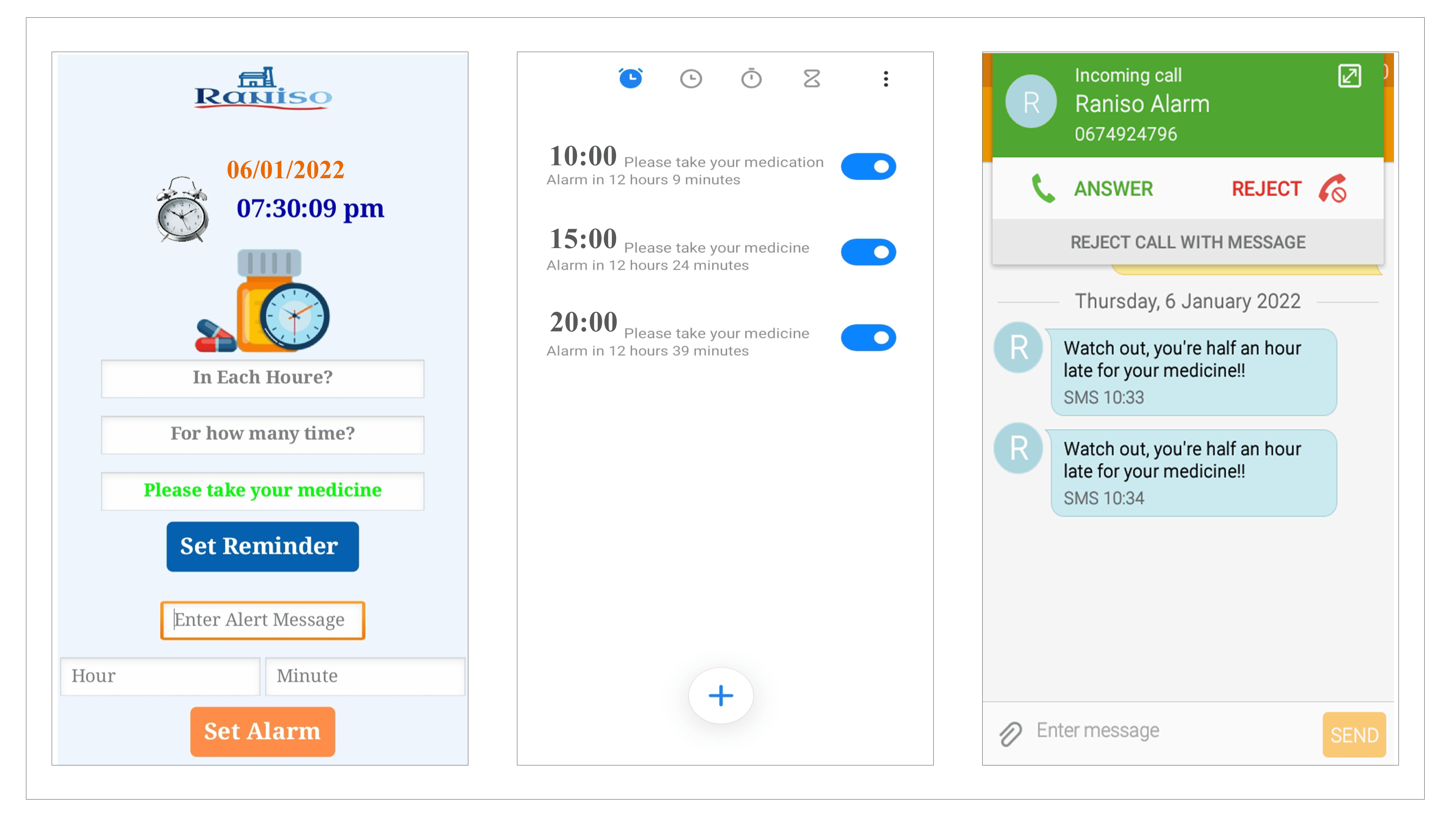}
		\caption{Smart medicine reminder interface on the Raniso.}
		\label{fig22_1}
	\end{figure}

This system includes a GSM module, DS3231 real-time clock module, LCD screen, push buttons, speaker, and LEDs. A speaker and LEDs are used to alert and remind that it’s time to take medicine. The LCD screen is set to cycle in three screens. The 1st screen shows the message “ Please take care of your health ”. The second screen is a help screen that tells to press the select push button to select any one-time slot to remind (once/twice/thrice in a day). The time slot is changeable in the program and can be configured accordingly. Four push buttons are used and each has a distinct select feature. The first push-button is used for reminding to take the medication once a day, the second is used to recall twice a day, and the third to recall three times a day. Stopping the notifications is done with the fourth push button when the user has taken his/her medication. If the patient is half an hour late in taking their medicine, the GSM module sends calls and short messages to alert them to hurry up to take their medication. In addition, the user can use the Raniso app’s smart medication reminder interface illustrated in Figure~\ref{fig22_1}, which will help them take their medicine on time, according to their treatment plan, and allow them to remotely manage and control medications/pill schedules and usage data. It also allows direct communication anytime anywhere between patients and caregivers, as it quickly alerts the caregiver if the patient needs help. The proposed IoT/M2M smart medication reminder system is extremely versatile and can be used by patients at home, doctors in hospitals, and many other places.

\subsection{Indoor Air Quality Monitoring System}

\begin{figure}[!h]
		\centering 
		\includegraphics[scale=0.4]{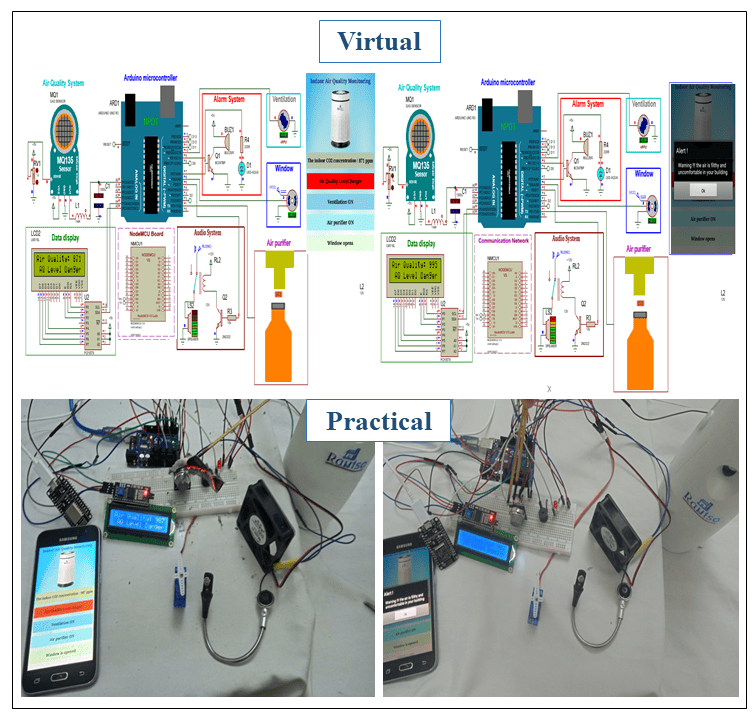}
		\caption{Indoor air quality monitoring system.}
		\label{fig23}
	\end{figure}

Indoor air pollution has become a real-life problem and a common phenomenon in buildings. With the spread of COVID-19, people spend most of their time indoors, and poor indoor air quality (IAQ) is a significant public health risk, which can increase short-term health problems such as fatigue and nausea, as well as chronic respiratory disease, heart disease, and lung cancer. It is now necessary to monitor air quality in real-time in most buildings. The air quality monitoring system developed in this work is based on the MQ135 sensor, DHT11, servo motor, DC fan, air purifier, speaker, buzzer, LCD screen, and LEDs (see Figure~\ref{fig23})).

\begin{figure}[!h]
		\centering 
		\includegraphics[scale=0.175]{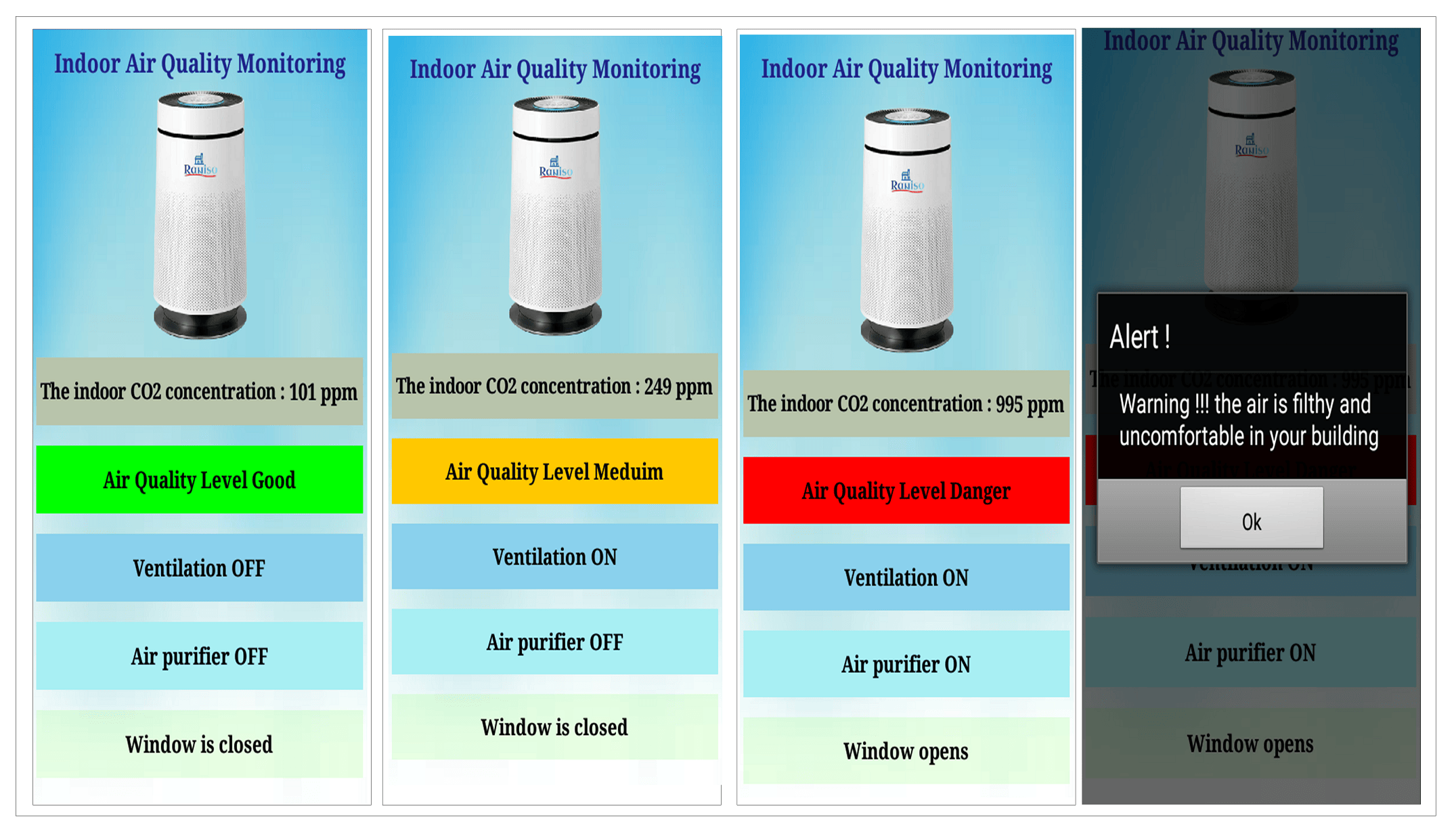}
		\caption{Indoor air quality monitoring interface on the Raniso.}
		\label{fig24}
	\end{figure}

\begin{figure}[!h]
		\centering 
		\includegraphics[scale=0.7]{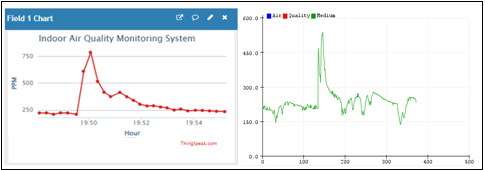}
		\caption{Indoor air quality monitoring graph on ThingSpeak and the Serial Plotter tool.}
		\label{fig25}
	\end{figure}

{IAQ is measured in Parts Per Million (PPM) units, where a lower PPM value indicates good air quality and a higher value indicates polluted air containing toxic gases.} When the value is less than 130 PPM, then the LCD and the Raniso App will display ”Air Quality Level Good ”. The fan will turn on when the value is between 130 PPM and 250 PPM. The LCD and the Raniso App will display ”Air Quality Level Medium. ” Whenever the value increases to 250 PPM. The buzzer will start beeping, the red LED will light up, the fan and the air purifier will turn on, and the LCD and the Raniso App will display” Air Quality Level Danger”. This service is connected to the Internet, and as a result, anyone can remotely visualize the air quality index from anywhere via the interface of air quality monitoring in the Raniso App (see Figure~\ref{fig24})). {This system aims to perform real-time monitoring of IAQ parameters via our channel in the ThingSpeak platform (see Figure}~\ref{fig25}) {and generate alerts via Raniso App to the building occupants to avoid hazardous conditions.}

\section{Results and Discussion}	
This section presents the findings from the prototype’s functional testing of the proposed IoT/M2M smart building system, which consists of several innovative services and functionalities such as smart parking, garden irrigation automation, intrusion alarm, smart door, fire and gas detection, smart lighting, smart medication reminder, and indoor air quality monitoring. We tried these services individually in the previous section, {while in this section, the effectiveness of the designed system is confirmed by testing all its services and functions on the final IoT / M2M smart building model}  {that was created to elaborate the performance and functionality of the proposed approach.}

 \begin{figure}[!h]
		\centering 
		\includegraphics[scale=0.25]{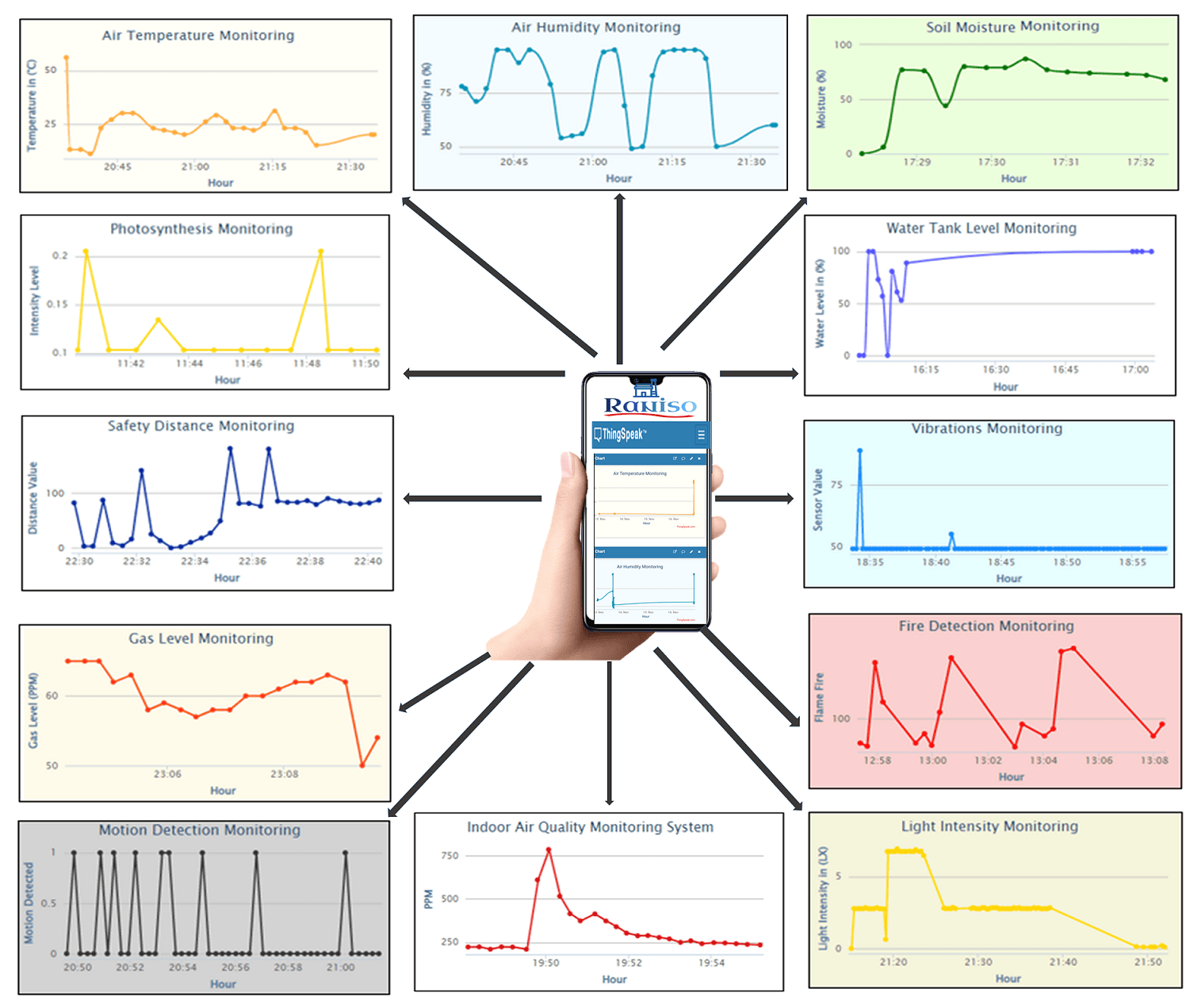}
		\caption{Quantitative sensing results for all the proposed applications.}
		\label{fig26}
	\end{figure}

Many sensors, actuators, and shields used resulted in a low power supply and a limited number of ports. As a result, we added the Arduino Mega board along with the Arduino UNO and NodeMCU for perfect working operation and fulfilled the requirements of the IoT/M2M smart building. Our goal was to design and implement an IoT/M2M smart building based on the convergence of two heterogeneous networks: wireless sensor networks operating on RFID, Bluetooth, Wi-Fi, and mobile cellular networks like GSM, 4G, or 5G. {All data collected from the suggested system is stored in the cloud database and sent to the ThingSpeak server and was visualized in four channels. The channels receive data from the building sensors at an interval of 10 seconds and are visualized as line graphs on the ThingSpeak webpage and the Raniso App in real-time.} Figure~\ref{fig26} {shows the graphs that were obtained from our channels in the ThingSpeak platform. In addition to the fact that these graphs show the possibility of monitoring building data in real-time anywhere via the Internet, it also shows the proposed approach's good performance and fast response. For example, garden irrigation automation graphs show how quickly the system responds in seconds should a plant need to be watered, photosynthesized, or refilled a water tank, etc. The same applies to other services, as the indoor air quality, fire, and gas detection graphs also show the system's speed in mitigating fire and gas leak risks. In addition to the automatic lighting service, whose graphs show real-time brightness, lux, and movement values.} 
The results we have reached are very satisfactory, as we have achieved the desired goal of providing accurate readings for the server. Besides, the latency is very low, so the users can easily control and monitor the smart building remotely anytime and anywhere using mobile phones through Raniso App. Also, one of this study’s accomplishments is using artificial intelligence to control appliances by allowing Google Assistant voice commands. The proposed smart building is customizable according to the user’s preferences. For example, the user can control the lights via pushbuttons, Bluetooth, Wi-Fi, 4G/5G networks, a voice through Google Assistant, or automatically based on the PIR sensor’s feedback. {Also, the proposed smart door, in which the user can control the keyless door, either by RFID card, password, or remotely via the Raniso app using the Bluetooth connection or the Internet via WiFi, 4G/5G.} The same for the rest of the proposed applications (smart parking, smart garden, etc.). {Furthermore, to examine the performance metric of the proposed system's error rate, and based on the data collected in }Table IV, {including temperature, humidity, and indoor air quality for 15 days, we were able to calculate the proposed system's error rate, which was less than 1\%.}
Table~\ref{data} presents the data collected, including the Temperature observed by the proposed system ($T_{ps}$), Temperature observed from Official Data ($T_{od}$), Humidity observed by the proposed system ($H_{ps}$), Humidity Data observed from Official Data ($H_{od}$), IAQ Data observed by the proposed system ($IAQ_{ps}$), and IAQ Data observed from Official Data ($IAQ_{od}$).  The data collected was taken over 15 days, which represents the number of observations (N). 

\begin{table}
 	\scriptsize
 	\renewcommand{\arraystretch}{1}
 	\caption{\uppercase{COLLECTED DATA OVER 15 DAYS.}}
 	\label{data}
 	\centering
 	\begin{tabular}{|c|c|c|c|c|c|c|}
 		\hline 
 		Date&$T_{ps}$&$T_{od}$&$H_{ps}$&$H_{od}$&$IAQ_{ps}$&$IAQ_{od}$ \\
    		\hline
 		01/11/2022&24$^{\circ}$C&25$^{\circ}$C&90\%&89\%&240&239 \\
 		\hline
            02/11/2022&27$^{\circ}$C&26$^{\circ}$C&76\%&78\%&225&228 \\
 		\hline
            03/11/2022&26$^{\circ}$C&25$^{\circ}$C&84\%&83\%&279&275 \\
 		\hline
            04/11/2022&23$^{\circ}$C&24$^{\circ}$C&77\%&78\%&459&470 \\
 		\hline
            05/11/2022&17$^{\circ}$C&19$^{\circ}$C&75\%&77\%&760&757 \\
 		\hline
            06/11/2022&18$^{\circ}$C&18$^{\circ}$C&77\%&76\%&500&499 \\
 		\hline
            07/11/2022&22$^{\circ}$C&23$^{\circ}$C&49\%&51\%&447&450 \\
 		\hline
            08/11/2022&23$^{\circ}$C&24$^{\circ}$C&50\%&53\%&330&333 \\
 		\hline
            9/11/2022&26$^{\circ}$C&25$^{\circ}$C&54\%&55\%&109&112 \\
 		\hline\
             10/11/2022&24$^{\circ}$C&24$^{\circ}$C&56\%&56\%&83&85 \\
 		\hline
            11/11/2022&24$^{\circ}$C&25$^{\circ}$C&47\%&46\%&100&96 \\
 		\hline
            12/11/2022&19$^{\circ}$C&20$^{\circ}$C&61\%&59\%&69&71 \\
 		\hline
            13/11/2022&21$^{\circ}$C&20$^{\circ}$C&78\%&78\%&168&163 \\
 		\hline
            14/11/2022&20$^{\circ}$C&19$^{\circ}$C&80\%&82\%&251&249 \\
 		\hline
             15/11/2022&19$^{\circ}$C&19$^{\circ}$C&99\%&98\%&270&266 \\
 		\hline
           	
	\end{tabular}
 \end{table}
For the temperature, the mean of $T_{ps}$ is given as
\begin{equation}\label{MT}
\begin{split}
       MT_{ps} =  \frac{\sum_{ i=1}^{N=15}T_{ps}}{N}.
       \end{split}
\end{equation}
The mean of $T_{od}$  is given as 
\begin{equation}\label{od}
\begin{split}
       MT_{od} =  \frac{\sum_{ i=1}^{N=15}T_{od}}{N},
       \end{split}
\end{equation} and the temperature error rate $e_{T}$  is expressed as 
\begin{equation}\label{et}
\begin{split}
       e_{T} =  \frac{ MT_{od}-MT_{ps}}{MT_{od}}*100.
       \end{split}
\end{equation}
For humidity, the mean of  $H_{ps}$  is given as
\begin{equation}\label{MH}
\begin{split}
       MH_{ps} =  \frac{\sum_{ i=1}^{N=15}H_{ps}}{N}.
       \end{split}
\end{equation}
The mean of $H_{od}$ is given as
\begin{equation}\label{oH}
\begin{split}
       MH_{od} =  \frac{\sum_{ i=1}^{N=15}H_{od}}{N},
       \end{split}
\end{equation}
and the (humidity error rate) $e_{H}$ is expressed as
\begin{equation}\label{eH}
\begin{split}
       e_{H} =  \frac{ MH_{od}-MH_{ps}}{MH_{od}}*100.
       \end{split}
\end{equation}
Similarly, for indoor air quality, the mean of  $IAQ_{ps}$  is given by 
\begin{equation}\label{M1}
\begin{split}
       MIAQ_{ps} =  \frac{\sum_{ i=1}^{N=15}IAQ_{ps}}{N}.
       \end{split}
\end{equation}
The mean of $IAQ_{od}$ is given by 
\begin{equation}\label{o1}
\begin{split}
       MIAQ_{od} =  \frac{\sum_{ i=1}^{N=15}IAQ_{od}}{N},
       \end{split}
\end{equation}
and the error percentage  is expressed as
\begin{equation}\label{e1}
\begin{split}
       e_{IAQ} =  \frac{ MIAQ_{od}-MIAQ_{ps}}{MIAQ_{od}}*100.
       \end{split}
\end{equation}
Based on the data collected in Table~\ref{data} and the mathematical equations from ~\eqref{MT} to ~\eqref{e1}, we were able to calculate the estimated error rate for temperature as 0.89\%, humidity as 0.56\%, and indoor air quality as 0.06\%. These error rates are very low, indicating the good performance and efficiency of the proposed system.
    
\section{Conclusion}
This paper focuses on developing an IoT/M2M smart building paradigm based on the convergence of wireless sensor networks and mobile-cellular networks. The proposed system used open hardware (Arduino, NodeMCU, sensors, modules, etc.) and open-source software (IDE, Proteus, and Raniso App). Based on our findings in this paper, we prove that our proposed solution offers a novel architectural design for a low-cost and flexible system that can be deployed for various smart IoT/M2M systems, including smart grids, smart retail, smart cities, etc. As an example, the architectural design was provided in more detail for the case of a smart building, where we suggested several main services and functionalities like smart parking, garden irrigation automation, intrusion alarm, smart door, fire and gas detection, smart lighting, smart medication reminder, and indoor air quality monitoring. All these services can be controlled and monitored remotely {via our channels in the ThingSpeak platform and through our multiple-platform mobile application named} ‘Raniso’, a local server that allows remote building control via RFID/Bluetooth/Wi-Fi connectivity and cellular networks like GSM, 4G, or 5G. The proposed IoT/M2M smart building was designed, implemented, deployed, and tested and yielded the expected results. This smart building system can be developed for future research purposes by integrating machine learning techniques to make it more robust and advanced.

\bibliographystyle{IEEEtran}
\bibliography{ref}
\end{document}